\documentclass[3p]{elsarticle}

\usepackage{amssymb}
\usepackage{amsmath}
\usepackage{xspace}
\usepackage{algorithm}
\usepackage{algorithmic}
\usepackage{textcomp} 
\usepackage{subfig}

\usepackage[
    dvipdfm,
    pdftitle={PetRBF---A parallel O(N) algorithm for radial basis function interpolation},
    pdfauthor={Rio Yokota, L.~A. Barba, Matthew G.~Knepley},
    pdfpagemode={UseOutlines},
    bookmarks, bookmarksopen,bookmarksnumbered={True},
    colorlinks, linkcolor={blue},citecolor={blue},urlcolor={blue}
    ]{hyperref}

\makeatletter
\def\url@leostyle{%
  \@ifundefined{selectfont}{\def\UrlFont{\sf}}{\def\UrlFont{\small\ttfamily}}}
\makeatother
\newcommand{\petsc}{\textsc{pets}c\xspace}

\newcommand{\pdes}{\textsc{pde}s\xspace}
\newcommand{\rbf}{\textsc{rbf}\xspace}
\newcommand{\rbfs}{\textsc{rbf}s\xspace}
\newcommand{\ddm}{\textsc{ddm}\xspace}
\newcommand{\ddms}{\textsc{ddm}s\xspace}
\newcommand{\fmm}{\textsc{fmm}\xspace}
\newcommand{\blas}{\textsc{blas}\xspace}
\newcommand{\lapack}{\textsc{lapack}\xspace}
\newcommand{\mpi}{\textsc{mpi}\xspace}
\newcommand{\ksp}{\textsc{ksp}\xspace}
\newcommand{\rasm}{\textsc{rasm}\xspace}
\newcommand{\gmres}{\textsc{gmres}\xspace}
\newcommand{\hector}{\textsc{hect}o\textsc{r}\xspace}
\newcommand{\bigON}{$\mathcal{O}(N)$}
\newcommand{\bigOsq}{$\mathcal{O}(N^2)$}

\begin{document}

\begin{frontmatter}



\title{PetRBF---A parallel \bigON\ algorithm for radial basis function interpolation}


\author[rio]{Rio Yokota}
\ead{Rio.Yokota@bristol.ac.uk}

\author[lab]{L.~A.~Barba \corref{lab2}  }
\ead{labarba@bu.edu}

\author[mgk]{Matthew G. Knepley}
\ead{knepley@ci.uchicago.edu}

\address[rio]{Department of Mathematics, University of Bristol, England BS8~1TW}
\address[lab]{Department of Mechanical Engineering, Boston University, Boston MA 02215}
\cortext[lab2]{Correspondence to:  110 Cummington St, Boston MA 02215, (617) 353-3883, \href{mailto:labarba@bu.edu}{labarba@bu.edu}}
\address[mgk]{Computation Institute, University of Chicago}

\begin{abstract}
We have developed a parallel algorithm for radial basis function (\rbf) interpolation that exhibits \bigON\ complexity,
requires \bigON\ storage, and scales excellently up to a thousand processes. The algorithm uses a \gmres iterative
solver with a restricted additive Schwarz method (\rasm) as a preconditioner and a fast matrix-vector
algorithm. Previous fast \rbf methods\,---\,achieving at most $\mathcal{O}(N\log N)$ complexity\,---\,were developed
using multiquadric and polyharmonic basis functions. In contrast, the present method uses Gaussians with a small
variance (a common choice in particle methods for fluid simulation, our main target application). The fast decay of the Gaussian basis function allows rapid convergence of the iterative solver even when the subdomains
in the \rasm are very small. The present method was implemented in parallel using the \petsc library (developer
version).  Numerical experiments  demonstrate its capability in problems of \rbf interpolation with more than $50$
million data points, timing at $106$ seconds ($19$ iterations for an error tolerance of $10^{-15}$) on 1024 processors
of a Blue Gene/L (700 MHz PowerPC processors).  The parallel code is freely available in the open-source model.
\end{abstract}

\end{frontmatter}


\section{Introduction}
\label{intro}



There are innumerable applications in computational science where one needs to perform approximation of a function based on finite data.  When the data are in a certain sense ``scattered'' in their domain, one very powerful technique is radial basis function (\rbf) interpolation.  Since the early 1980s, after a comparative study of various methods for scattered-data interpolation showed the superiority of \rbf methods \cite{franke1982}, a large amount of interest and research effort has been invested in this subject.  For many years, however, the wide applicability of \rbf interpolation was hindered by their numerical difficulty and expense.  Indeed, in their mathematical expression, \rbf methods produce a linear system of size equal to the number of data points.  A direct solution of such systems\,---\,requiring $\mathcal{O}(N^3)$ operations and \bigOsq\ memory usage\,---\,becomes prohibitive for more than a few thousand data points.  Great progress has been made in recent years towards alleviating this computational burden.  We review this progress below. The present work aims to position itself as the fastest and most efficient algorithm to date for \rbf interpolation.  It moreover exhibits excellent parallel scalability, and we put it to the test with up to 50 million unknowns.

In addition to scattered-data interpolation, \rbf methods are now included in the broad category of meshfree methods for the solution of partial differential equations.  The conventional methods for solving \pdes rely on a discretization of the domain in terms of a mesh, \emph{i.e.}, a set of inter-connected nodes.  In meshfree methods, the aim is to provide a means of discretization that relies only on a set of nodes, but with no knowledge of the connectivity among them.  One of the motivations is that the generation of a mesh (for example, as used in finite difference and finite element methods) can be a cumbersome task, especially for problems with complex geometry and moving boundaries. It is often cited that mesh generation can take up to 80\% of the time required for computational analysis in industrial applications \cite{HughesETal2005}. Indeed, this is the \emph{greatest hurdle }for the wide use of computational mechanics as a design tool.

Meshfree methods for the solution of \pdes can be grouped into two broad categories. One, as already mentioned, includes methods based on \rbf interpolation. The second is based on the least squares technique. To this second class of methods belong element-free Galerkin methods \cite{BelytschkoETal94}, reproducing kernel particle methods \cite{KamLiuETal1995}, $hp$-clouds methods \cite{DuarteOden96}, partition of unity methods \cite{babuska1997}, and meshless local Petrov-Galerkin methods \cite{atluri1998}.  We will not refer to least squares methods further.  They differ from \rbf methods in that they do not satisfy the interpolation conditions at the data. In certain applications, \emph{e.g.}, when it is known that the data contain noise, the smoothing quality of least-squares methods may be desirable.  We restrict this work to the interpolating \rbf methods, which have superior approximation quality without smoothing.

\rbf interpolation has a number of favorable characteristics. Madych \textit{et al.} \cite{madych+nelson1990} have shown that multiquadric \rbf interpolation schemes converge faster as the dimension increases, and converge exponentially as the density of the nodes increases. These favorable characteristics encouraged others to use it as a tool for solving \pdes. In the early 1990s, Kansa \cite{kansa1990a,kansa1990b} used \rbf interpolation in a global collocation method with multiquadrics to solve \pdes. He found that  \rbfs could yield a very accurate solution for parabolic and elliptic \pdes. This method has been further extended to symmetric collocation by Fasshauer \cite{fasshauer1997}, to modified collocation by Chen \cite{chen2002a} and to indirect collocation by Mai-Duy \cite{mai-duy2003}.

Amongst the \emph{un}favorable characteristics of \rbf interpolation, the most severe is the ill-conditioning of the linear system it produces. Preconditioning can result in significant clustering of eigenvalues and improves the condition numbers of the interpolation problem by several orders of magnitude \cite{beatsonETal1999}. Considerable progress in the iterative solution of multiquadric and polyharmonic \rbf interpolation was made with the development of preconditioners using approximate cardinal functions based on a subset of the interpolating points. This approach was first proposed by Beatson and Powell \cite{beatson1993}. Beatson \textit{et al.} \cite{beatsonETal1999} coupled this method with the \gmres iterative solver and a fast matrix-vector algorithm for polyharmonic splines to construct a method with $\mathcal{O}(N)$ storage and $\mathcal{O}(N\log N)$ complexity. Ling and Kansa \cite{ling2005} showed that an approximate least squares cardinal function preconditioner can be used to transform an ill-conditioned system arising from \rbf solution of \pdes into a well-conditioned system. Faul \textit{et al.} \cite{faul+goodsell+powell2005} used a carefully selected set of $q$ points to construct a preconditioner for which they were able to interpolate problems with $d=40$ dimensions and $N=10,000$ points in a few iterations. Gumerov and Duraiswami \cite{gumerov+duraiswami2007} used the fast multipole method (\fmm) to accelerate the matrix-vector multiplication in Faul's method and reduced the calculation cost from $\mathcal{O}(N^2)$ to $\mathcal{O}(N\log N)$. There have been a few other studies where fast summation methods were used to accelerate the matrix-vector multiplications for polyharmonic \cite{beatson+newsam1992,powell1993,gumerov2005}, multiquadric \cite{cherrie+beatson+newsam2002,roussos2005}, and Gaussian \cite{greengard+strain1991,roussos2005} basis functions.

When handling problems with large numbers of data points (say, millions of points), the large amount of memory usage can also become a problem. As the problem size grows, parallelization on distributed memory architectures becomes necessary. Therefore, domain decomposition methods (\ddm) \cite{smith1996} have been gaining interest. The idea behind \ddm is to divide the considered domain into a number of subdomains and then try to solve the original problem as a series of subproblems that interact through the interfaces. Hardy \cite{hardy1990} was the first to use \ddms for \rbf interpolation. Beatson \textit{et al.} \cite{beatson.et.al2000} used the multiplicative Schwarz method along with a coarse-grid preconditioner, and a fast matrix-vector algorithm for polyharmonic splines. Their code requires $\mathcal{O}(N)$ storage and $\mathcal{O}(N\log N)$ time, and the timings are quite impressive. There have been a few related efforts using the truncated multiquadric function \cite{kansa+hon2000}, multi-zone methods \cite{wong1999}, and a comparison between overlapping and non-overlapping \ddms for matching and non-matching subdomain interfaces \cite{li2004}. Ling and Kansa \cite{ling2004} used the additive Schwarz method and found that increasing the number of subdomains not only reduces the operation count but also increases the rate of convergence. However, none of these studies mention the parallelizability of these potentially highly-parallel methods, as they all view the \ddms merely as a way to deal with the ill-conditioning of the system.

All work with \ddms mentioned above involved global and conditionally positive-definite \rbfs, such as multiquadrics and thin-plate splines. For Gaussian basis functions, the choice and effectiveness of the preconditioner, \ddm, and fast matrix-vector algorithm differ substantially. In fact, for Gaussian basis functions with a small standard deviation $\sigma$, the resulting system is not so ill-conditioned, and preconditioning becomes a straightforward task of truncating the effect of the far field \cite{BarbaRossi2009}. In such a case, \ddms will provide this function of preconditioning by localizing the interaction, \textit{i.e.}, making the bandwidth of the matrix smaller. Furthermore, the matrix-vector multiplication could be performed by a fast Gauss transform \cite{greengard+strain1991} if $\sigma$ were much larger than the inter-node spacing $h$. However, for problems that have $\sigma\approx h$, it is faster to perform a direct calculation for the neighboring nodes. These specializations for the domain decomposition strategy and matrix-vector multiplication have a large positive impact on the performance and, as we show in this work, result in a method which has $\mathcal{O}(N)$ storage and $\mathcal{O}(N)$ time.

An application that will benefit tremendously from fast \rbf interpolation using Gaussian basis functions with small $\sigma$, is the vortex particle method. It is now a well-known fact that particle methods for continuum systems generally require some sort of spatial adaptation, and \rbf interpolation can be used in a number of different ways to handle this. The use of \rbf interpolation for spatial adaptation in the vortex particle method was proposed and demonstrated in \cite{barba2005}. It was shown that the core-spreading method \cite{leonard1980} coupled with \rbf interpolation was more accurate than the standard particle strength exchange scheme with remeshing. A parallel version of the vortex particle method with \rbf interpolation for spatial adaptation was later developed using the \petsc library \cite{petsc-manual}, \cite{barbaETal2005}. The parallel vortex code with \rbf interpolation was later used in a comprehensive study of vortex tripole evolution \cite{barba+leonard2007}. The same method was also used in a calculation of homogeneous isotropic turbulence in three dimensions, where the results of the vortex particle method agreed quantitatively with those of the reference calculation using a pseudo-spectral method with the same number of calculation points \cite{yokota2007}. However, the calculation time for the \rbf interpolation became dominant, compared to the other parts of the code. Although convergence of the Krylov method was generally observed with the parameters chosen in the bulk of experiments that were performed, no investigation was pursued at that time on the conditioning and convergence of the \rbf interpolation for this particular type of basis function. The generality of \rbf interpolation allows the possibility to extend this method to other meshfree methods\,---\,typically the ones that involve convection of the nodes, for which the basis functions must have asymptotically local influence (like the Gaussian).

The main goal of this paper is to develop a highly-parallel \rbf interpolation algorithm by using \gmres with the restricted additive Schwarz method (\rasm) \cite{cai1997} as a preconditioner along with a fast matrix-vector multiplication. We demonstrate that, for Gaussian basis functions with small $\sigma$, it is possible to eliminate global communications completely by exploiting the asymptotic locality of the basis function. This results in $\mathcal{O}(N/N_{procs})$ storage and close to $\mathcal{O}(N/N_{procs})$ complexity, where $N_{procs}$ is the number of processes. A serial version of the present method was investigated by Torres and Barba \cite{TorresBarba2009}. Their results show excellent convergence rates for different types of data, but they only show a limited portion of what this method truly has to offer. Unlike the previous studies on domain decomposition-based \rbf interpolation, our focus is on the optimum subdivision of the domain, and parallel scalability.

\section{RBF interpolation with domain decomposition}
\label{method}


\subsection{Background on radial basis function interpolation}

The technique of radial basis function (\rbf) interpolation was introduced as a tool for solving multivariate scattered data interpolation problems.  There has been a large production of research results in this field, with some excellent books being published in recent years \cite{buhmann2003,fasshauer2007}. 

The problem of scattered data interpolation is that of approximating a function $f$ at an arbitrary point $\vec{x}$, when its values are only known on a limited set of points $\vec{x}_j\in\mathbb{R}^d,\ j=1,...,N$. This is done by representing the function $f$ by the interpolant $s(\vec{x})$:
\begin{equation}
s(\vec{x}) = \sum_{j=1}^N  \lambda_j\  \phi(||\vec{x}-\vec{x}_j||)+\sum_{k=1}^M\gamma_k\ p_k(\vec{x}),
\label{eq:rbf1}
\end{equation}

\noindent where $\phi$ is the radial basis function, $||\cdot||$ denotes the Euclidean norm, and $\left\{p_k(\vec{x})\right\}_{k=1}^{M}$ is a basis for $\Pi_{M}^{d}$ (space of all $d$-variate polynomials with degree of less than $M$). Micchelli \cite{Micchelli1986} showed that there is a direct relationship between $\phi(r)$ being positive definite and the function $\phi(\sqrt{r})$ being completely monotone. If $\phi(\sqrt{r})$ is completely monotone, then the resulting system is positive definite on its own and does not need to be augmented by polynomials, which is the case for Gaussians and inverse multiquadrics. On the other hand, if $(-1)^M \phi^{(M)}(\sqrt{r})$ is completely monotone for some $M > 0$, then $\phi(r)$ is said to be \textit{conditionally} positive definite of order $M$, and requires augmentation by polynomials of degree $M-1$ to become positive definite.

The coefficients $\lambda_j$ and the polynomial functions $p_k(\vec{x})$ are chosen to satisfy the fitting conditions,
\begin{equation}
s(\vec{x}_j)=f(\vec{x}_j),\quad j=1,...,N
\end{equation}
along with the additional constraints
\begin{equation}
\sum_{j=1}^{N}\lambda_jp_k(\vec{x}_j)=0,\quad k=1,...,M
\end{equation}
This results in a linear system
\begin{equation}
\underbrace{
\left(
\begin{array}{cc}
\Phi & P\\
P^T & O
\end{array}
\right)
}_\mathcal{A}
\underbrace{
\left(
\begin{array}{c}
\vec{\lambda}\\
\vec{\gamma}
\end{array}
\right)
}_{\vec{x}}
=
\underbrace{
\left(
\begin{array}{c}
\vec{f}\\
0
\end{array}
\right)
}_{\vec{b}}
\label{eq:rbf2}
\end{equation}

\noindent where:
\begin{eqnarray*}
 \left\{\left\{\Phi_{ij}=\phi(||\vec{x}_i-\vec{x}_j||)\right\}_{i=1}^{N}\right\}_{j=1}^{N}
 \, , \,
 \left\{\left\{P_{ij}=p_j(\vec{x}_i)\right\}_{i=1}^{N}\right\}_{j=N+1}^{N+M}
  \, , \\
 \left\{\vec{\lambda}=\lambda_j\right\}_{j=1}^{N}
  \, , \,
 \left\{\vec{\gamma}=\gamma_j\right\}_{j=N+1}^{N+K}
 \, , \, \text{and} \,
 \left\{\vec{f}=f_i\right\}_{i=1}^{N}. 
\end{eqnarray*}

 For future reference of Equation~(\ref{eq:rbf2}) we denote the coefficient matrix as $\mathcal{A}$, the solution vector as $\vec{x}$, and the right hand side as $\vec{b}$. We use the symbol $O$ to represent the zero matrix of appropriate size, while the symbol 0 is used for the zero vector/scalar. Popular choices for the radial basis function $\phi$ are summarized in Table \ref{tab:rbf}.
\begin{table}
\caption{Examples of radial basis functions}
\label{tab:rbf}
\begin{center}
\begin{tabular}{ll}
\hline
Name of \rbf &\hspace{10mm} $\phi(r)$ \\
\hline
Gaussian&\hspace{10mm}$e^{-\epsilon^2 r^2}$\\
Polyharmonic&\hspace{10mm}
$
\left\{
\begin{array}{l}
r^\nu,\nu\in2\mathbb{N}-1\\
r^\nu\log r,\nu\in2\mathbb{N}
\end{array}
\right.
$
\\
Multiquadric&\hspace{10mm}$(1+\epsilon^2r^2)^\nu,\nu>0,\nu\notin\mathbb{N}$\\
Inverse\ multiquadric&\hspace{10mm}$(1+\epsilon^2r^2)^\nu,\nu<0$\\
\hline
\end{tabular}
\end{center}
\end{table}

\subsection{Restricted additive Schwarz method (\rasm)}\label{sse:rasm}

\begin{figure}
  \centering
  \includegraphics[keepaspectratio=true,width=0.5\textwidth]{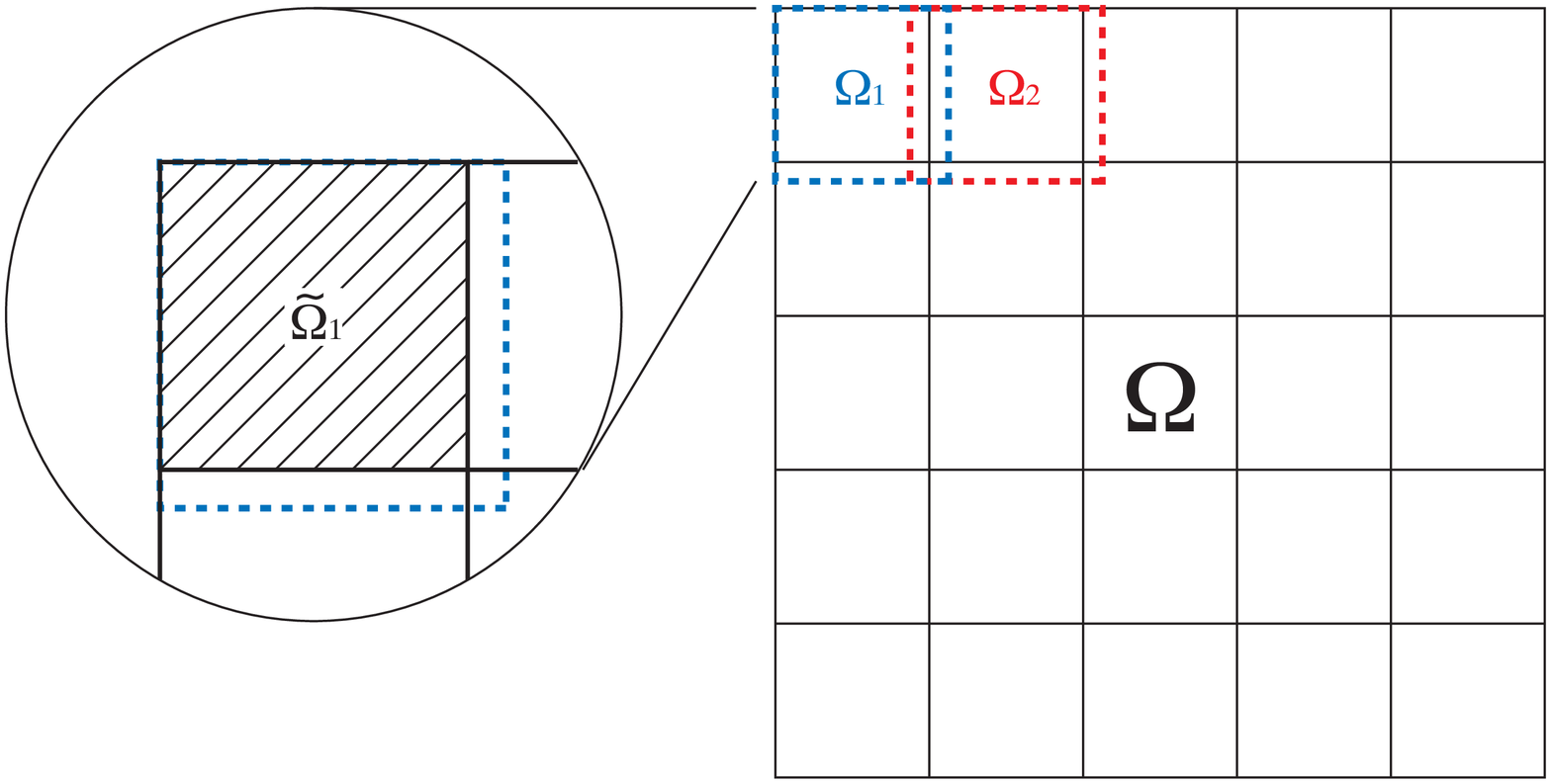}
  \caption{Illustration of Overlapping and Non-overlapping Domains}
  \label{fig:dd}
\end{figure}

The idea behind domain decomposition methods is to divide the considered domain into a number of subdomains and then try to solve the original problem as a series of subproblems that interact through the interfaces. A brief explanation is given below to introduce the main concepts of domain decomposition methods, for completeness. The key components that will be explained in this section are used to describe the parallelization procedure and analyze the results later on in this paper.

We consider a square domain $\Omega$ as shown in Figure \ref{fig:dd}, having a somewhat uniform distribution of calculation points within the domain. These calculation points could be the nodes of a finite difference/element mesh or they could be particles. Let us divide the entire calculation domain $\Omega$, into overlapping subdomains $\Omega_i$. Also, let $\tilde{\Omega}_i$ denote the non-overlapping subdomains as shown in Figure \ref{fig:dd}. The solution of a linear system $\mathcal{A}\vec{x}=\vec{b}$ for the whole domain can be obtained by sequentially solving the individual overlapping subdomains $\mathcal{A}_i\vec{x}_{\Omega_i}=\vec{b}_{\Omega_i}$, where $\mathcal{A}_i$, $\vec{x}_{\Omega_i}$, and $\vec{b}_{\Omega_i}$ are the sub-elements associated with the domain $\Omega_i$ for $\mathcal{A}$, $\vec{x}$, and $\vec{b}$, respectively. The values $\vec{x}$ in the overlapping region can be overwritten when the next subdomain is solved. If the values $\vec{x}$ of the previous subdomains are used to update the solution of the present subdomain calculation, it is called a \emph{multiplicative} Schwarz method. If each subdomain is solved individually and the solution of the entire domain is updated simultaneously at the end of each iteration step, it becomes an \emph{additive} Schwarz method. Furthermore, when the values $\vec{x}$ outside of the subdomain $\tilde{\Omega}_i$ are discarded after the calculation of each subdomain $\Omega_i$, it is called the \emph{restricted additive} Schwarz method (\rasm) \cite{cai1997}. The restricted additive Schwarz method is known to converge faster than the additive Schwarz method and requires less communication in parallel calculations \cite{efstathiou2003}.  As far as we know, \rasm has not been used before within \rbf\ interpolation with domain decomposition.

The mathematical formulae for the above methods can be written quite succinctly by defining a restriction matrix $R_i$,
\begin{equation}
\vec{x}_{\Omega_i}=R_i\vec{x}=(I \quad 0)\left(
\begin{array}{c}
\vec{x}_{\Omega_i}\\
\vec{x}_{\Omega\backslash\Omega_i}
\end{array}
\right)
\end{equation}

\noindent which restricts a vector in the whole domain $\vec{x}$ to one in the subdomain $\vec{x}_{\Omega_i}$. If the entire domain has $N$ elements and the subdomain has $N_i$ elements, $R_i$ is a $N_i\times N$ matrix that restricts the vector $\vec{x}$ of size $N$ into vector $\vec{x}_{\Omega_i}$ of size $N_i$. The transpose of the restriction matrix is the projection matrix,
\begin{equation}
\vec{x}=R_{i}^{T}\vec{x}_{\Omega_i}
\end{equation}

\noindent which projects the vector in the subdomain onto the whole domain. Note that the $R_i$ matrices are usually not formed in practice, but are introduced to express the different algorithms in a concise manner. Using the restriction and projection matrices, the multiplicative Schwarz method for the $n$-th iteration step with $p$ overlapping subdomains can be written as

\begin{align}
\vec{x}^{n+1/p}&=\vec{x}^{n}+R_{1}^{T}\mathcal{A}_{1}^{-1}R_1(\vec{b}-A\vec{x}^n)\nonumber\\
\vec{x}^{n+2/p}&=\vec{x}^{n+1/p}+R_{2}^{T}\mathcal{A}_{2}^{-1}R_2(\vec{b}-A\vec{x}^{n+1/p})\label{eq:msm}\\
&\dots\nonumber\\
\vec{x}^{n+1}&=\vec{x}^{n+(p-1)/p}+R_{p}^{T}\mathcal{A}_{p}^{-1}R_p(\vec{b}-A\vec{x}^{n+(p-1)/p})\nonumber
\end{align}

\noindent where $\mathcal{A}_i=R_i\mathcal{A}R_{i}^{T}$. The $n$-th iteration step with $p$ overlapping subdomains of the additive Schwarz can be expressed as
\begin{equation}
\vec{x}^{n+1}=\vec{x}^{n}+\sum_{i=1}^{p}R_{i}^{T}\mathcal{A}_{i}^{-1}R_i(\vec{b}-A\vec{x}^n)
\label{eq:asm}
\end{equation}

The only difference between the multiplicative Schwarz method in Equation~(\ref{eq:msm}) and the additive Schwarz method in Equation~(\ref{eq:asm}) is that the multiplicative Schwarz method updates the residual after each subdomain is calculated, whereas the additive Schwarz method updates the residual only once per iteration.

Equation~(\ref{eq:asm}) can be extended to the restricted additive Schwarz method (\rasm) by defining a projection matrix
\begin{equation}
\vec{x}=\tilde{R}_{i}^{T}\vec{x}_{\tilde{\Omega}_i}
\end{equation}
which projects only the values in the \textit{non-overlapping} subdomain $\vec{x}_{\tilde{\Omega}_i}$ onto the whole domain. The  $n$-th iteration step of the \rasm is
\begin{equation}
\vec{x}^{n+1}=\vec{x}^{n}+\sum_{i=1}^{p}\tilde{R}_{i}^{T}\mathcal{A}_{i}^{-1}R_i(\vec{b}-A\vec{x}^n)
\label{eq:rasm}
\end{equation}

\noindent where the matrix $R_i$ restricts the residual to the overlapping subdomain $\Omega_i$, $\mathcal{A}_{i}^{-1}$ solves the problem on $\Omega_i$, and $\tilde{R}_{i}^{T}$ projects the solution in the non-overlapping subdomain $\tilde{\Omega}_i$ onto the entire domain. The matrix
\begin{equation}
\mathcal{M}^{-1}=\displaystyle\sum_{i=1}^{p}\tilde{R}_{i}^{T}\mathcal{A}_{\Omega_i}^{-1}R_i
\end{equation}
can be viewed as a preconditioner, and can be used along with any Krylov subspace method. In the present calculation we use the \gmres \cite{saad1996}.

The calculation of inner products, norms, and scalar multiplication in the \gmres are not computationally intensive nor do they require large data transfer. Therefore, the efficiency and parallelism of the \gmres calculation depends on how the preconditioning is handled. When the preconditioning is done by \rasm, the efficiency of the calculation is strongly affected by the size of the overlapping and non-overlapping subdomains. One needs to find the balance between having a large number of small subdomains against having a small number of large subdomains. Also, increasing the overlap region allows the \gmres to converge in fewer iterations, but each iteration will take longer because the direct solve will be performed for larger overlapping subdomains. We have performed a thorough investigation of the optimum subdomain size for our algorithm, and present the results in \S\ref{serial}.

\subsection{Parallelization of \rasm}
\begin{figure}[tbp]
  \centering
  \subfloat[One subdomain per process]{
  \includegraphics[keepaspectratio=true,width=0.75\textwidth]{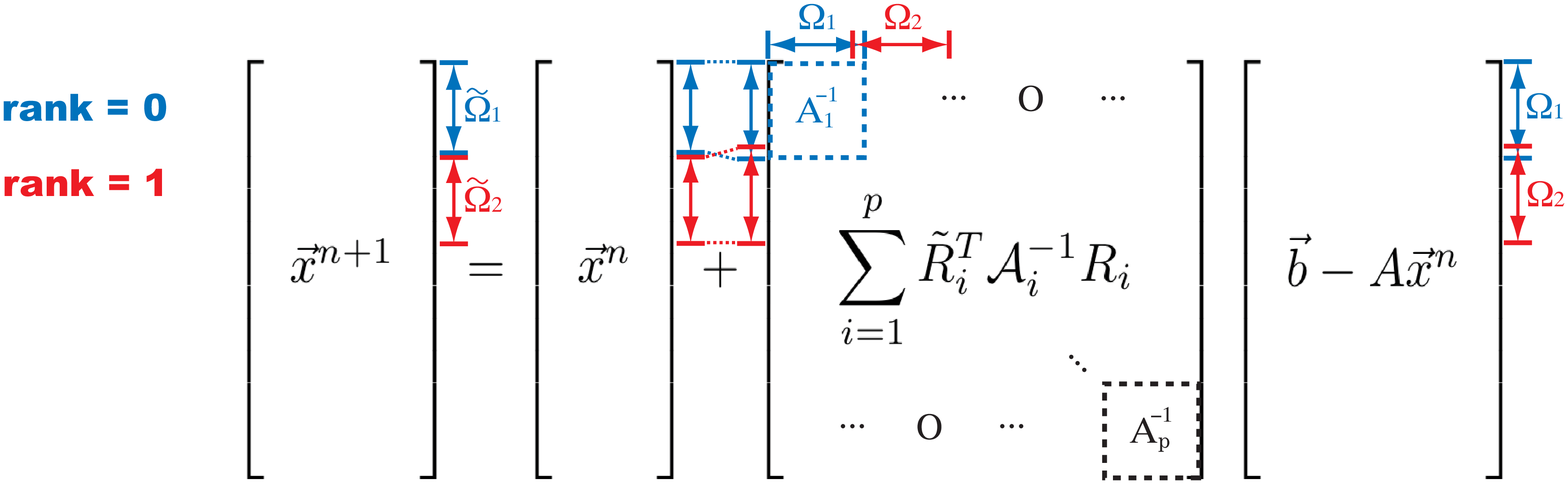}\label{fig:par}}\\
  \subfloat[Multiple subdomains per process]{
  \includegraphics[keepaspectratio=true,width=0.75\textwidth]{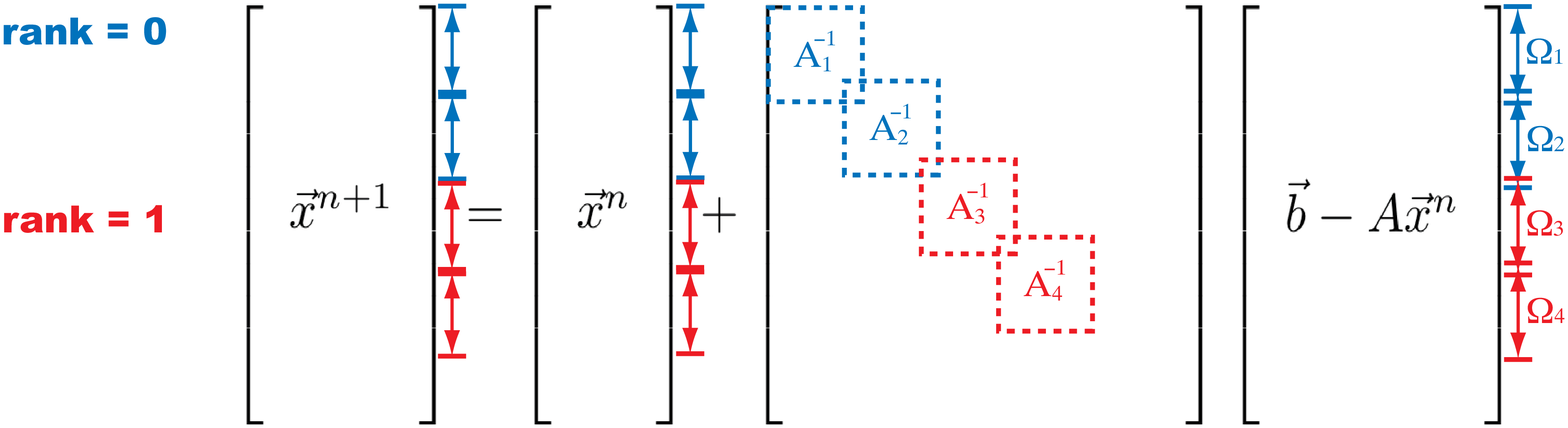}\label{fig:par2}}
  \caption{Parallelization strategy of the \rasm}
  \label{fig:para}
\end{figure}

For the parallelization, we assume a distributed memory model that stores only the local components of each vector. Parallelization of the calculation of inner products, norms, and scalar multiplication are trivial, and the only noteworthy task when parallelizing the preconditioned \gmres is the preconditioning itself. When the \rasm is used as the preconditioner, it can be seen that Equation~(\ref{eq:rasm}) can be readily parallelized by assigning each subdomain to a separate process. A schematic of the parallelization of  the \rasm corresponding to Equation~(\ref{eq:rasm}) is shown in Figure \ref{fig:para}\subref{fig:par}. The different colors represent separate processes. The ``rank'' refers to the rank of the MPI process. Each process owns a local portion of the global vectors $\vec{x}$ and $\vec{b}$. The partitioning of the vectors corresponds to the non-overlapping subdomains $\tilde{\Omega}_i$ shown in Figure \ref{fig:dd}. For the case of one subdomain per process, as seen in  Figure \ref{fig:para}\subref{fig:par}, each process owns one submatrix $\mathcal{A}_i$. Each process would require a small amount of communication from other processes to calculate $\mathcal{A}\vec{x}^n$. Once this is done, the local residual for the overlapping subdomain $\vec{r}_i=\vec{b}_i-\mathcal{A}\vec{x}_{i}^{n}$ can be used as the right-hand-side to solve the local system. Then, the non-overlapping part of the solution to the local system is added to the vector $\vec{x}^{n}$. This is done by simply updating only the local values that belong to the process. This procedure is depicted in Figure \ref{fig:para}\subref{fig:par}.

For \rbf interpolation using basis functions with negligible global effects, the matrix $\mathcal{A}$ can be considered to have a finite bandwidth. Thus, the calculation of $\mathcal{A}\vec{x}_{i}^{n}$ can also be done somewhat locally. Let us consider a normalized Gaussian basis function in 2D as an example. The matrix $\mathcal{A}$ has the following elements:
\begin{equation}
\mathcal{A}_{ij}=\frac{1}{2\pi\sigma^2}\exp\left(-\frac{|\mathbf{x}_i-\mathbf{x}_j|^2}{2\sigma^2}\right)
\end{equation}

Since the Gaussian function decays rapidly, the elements of matrix $\mathcal{A}$ corresponding to the interaction of distant points can be neglected. This ``sparsity'' of $\mathcal{A}$ depends on the relative size of the calculation domain compared to the standard deviation, $\sigma$, of the Gaussian function. For problems with relatively small $\sigma$, the calculation of $\mathcal{A}\vec{x}_{i}^{n}$ can be done by a sparse matrix-vector multiplication with controllable accuracy. If $\sigma$ is kept constant while the size of the calculation domain is increased along with $N$, the calculation load will scale as $\mathcal{O}(N)$. The communication required to perform $\mathcal{A}\vec{x}_{i}^{n}$ is also limited to a constant number of elements in the vicinity. Therefore, the \rasm becomes an extremely parallel algorithm with minimum communication for the \rbf interpolation using Gaussian basis functions with small $\sigma$.

In the present work, domain decomposition is used not only to save storage, but also to increase the speed of convergence. Therefore, it is useful to have independent control over the subdomain size and the number of processes, as shown in Figure \ref{fig:para}\subref{fig:par2}. In this case, each process stores multiple sets of $\vec{x}_i$ and $\vec{b}_i$, and loops over the the subdomain solves. The restriction from overlapping subdomains to the non-overlapping subdomains is not as straightforward as the case of one subdomain per process, because the local portion of the vector $\vec{x}$ does not necessarily correspond to the values that need to be updated. In the present algorithm, we perform the restriction by using a separate index list for the non-ovelapping subdomain.


\subsection{Implementation details}

The implementation of the algorithm developed here has been realized using the Portable, Extensible Toolkit for Scientific Computation (\petsc) \cite{petsc-manual}. This is a library developed over many years at Argonne National Laboratories\footnote{\href{http://www.mcs.anl.gov/petsc/}{http://www.mcs.anl.gov/petsc/}}. It is fine-tuned and well-supported, and provides an interface between scientific application codes and low-level libraries such as \blas, \lapack, and \mpi. All vectors and matrices are distributed among processes by \petsc and only the local portion is stored. Data types such as \texttt{Vec} and \texttt{Mat} allow the user to create these distributed objects and access them using global indices. The fact that all \mpi communications happen ``under the hood" allows users to manage parallel codes (almost) as if they were serial codes. Another feature of \petsc is that it can switch between different Krylov subspace solvers and preconditioners with just one command line option.

The current version of \petsc (3.0.0, released December 2008) provides routines for domain decomposition using additive Schwartz methods. The goal of these routines seems to have been using domain decomposition for parallelization only, and thus they were written to hold one subdomain per process.  To fit our needs, one of the authors (Matthew G. Knepley, who is a member of the \petsc developer team) has extended the \petsc functions  to handle multiple subdomains per process, and thus we are able to implement the restrictive additive Schwarz method as used in our algorithm. This extension to \petsc (currently only available in the developer version) enabled us to implement our \rbf interpolation algorithm with excellent parallel scalability.

\medskip

In our implementation, we use the following features of the \petsc library.
 \vspace{-0.5em}
\begin{itemize}
\item[$\triangleright$] vectors (for vectors $\vec{x}$ and $\vec{b}$) \vspace{-0.5em}
\item[$\triangleright$] ghost vectors (for overlapping subdomains and matrix-vector multiplication) \vspace{-0.5em}
\item[$\triangleright$] index sets (for setting overlapping and non-overlapping subdomains) \vspace{-0.5em}
\item[$\triangleright$] shell matrices (for matrices $\mathcal{A}$ and $\mathcal{M}$) \vspace{-0.5em}
\item[$\triangleright$] \ksp linear solvers (\gmres, \rasm)
\end{itemize}
The parallel \gmres-\rasm method that has been described earlier in this section can be implemented in \petsc using the above features. The vectors are distributed among the processes and the local storage is $N/N_{proc}$ for each vector, where $N$ is the number of elements and $N_{proc}$ is the number of processes. The calculation of inner products, norms, and scalar multiplication are done by calling \petsc routines. The ghost vectors handle the non-local values that are necessary for the direct solve for the overlapping subdomains and also the matrix vector multiplication of $\mathcal{A}\vec{x}$. One scatters the global vector into a local ``ghosted" work vector, performs the computation on the local work vectors, and then scatters back into the global solution vector. The index sets are global indices that are used to define the elements in the overlapping and non-overlapping subdomain. Although their numbering is global, the index sets are distributed among the processes in the same way as the vectors. The shell matrices allow one to calculate the matrix vector multiplication $\mathcal{A}\vec{x}$ without actually storing the matrix $\mathcal{A}$. For the preconditioner $\mathcal{M}$, only the small matrices for the subdomain $\mathcal{A}_i$ are formed. The matrix $\mathcal{A}_i$ is then used to calculate Equation~(\ref{eq:rasm}). After all these objects are defined, \petsc finally calculates the linear system solution using the \texttt{KSPSolve()} command, which takes the solver type \gmres (default) and the preconditioner type \rasm from the command line options.

\section{Parametric study of convergence rate}
\label{serial}


The rate of convergence, calculation time, storage requirements, and communication requirements of the present method depend strongly on five independent parameters. Figure \ref{fig:param} illustrates these parameters as they refer to the domain decomposition choices, and basis function choices. In this section, we present the results of a non-dimensionalized parametric study to demonstrate the sensitivity of the convergence rate and calculation time to these parameters.

All the calculations presented here were obtained on the Cray XT4 machine on the \hector system at the UK National Supercomputing Service \footnote{\href{http://www.hector.ac.uk/}{http://www.hector.ac.uk/}}. \hector offers a total of 5664 \textsl{AMD} 2.3 GHz quad-core \textsl{Opteron} processors. The system offers 8 GB of memory per processor, which is shared between its four cores. The processors are connected with a high-bandwidth interconnect using \textsl {Cray SeaStar2} communication chips. Only \emph{one processor} of this machine was used for obtaining the results in this section\,---\,parallel results are presented in the next section.

\subsection{Calculation conditions}

\begin{figure}
  \centering
  \includegraphics[keepaspectratio=true,width=0.7\textwidth]{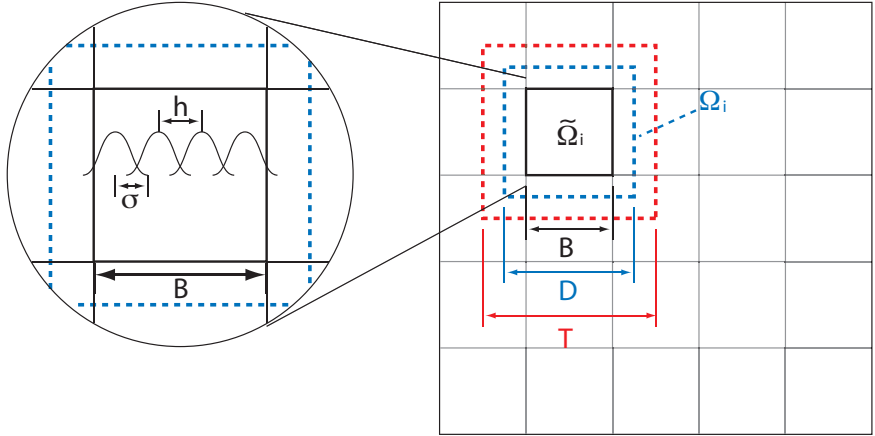}
  \caption{Illustration of parameters in the \rasm for \rbf interpolation using Gaussian basis functions}
  \label{fig:param}
\end{figure}

The whole domain is divided into non-overlapping subdomains $\tilde{\Omega}_i$ and overlapping subdomains $\Omega_i$ as shown in Figure \ref{fig:param}. We define $B$ as the size of the non-overlapping subdomain and $D$ as the size of the overlapping subdomain. We also define an additional subdomain with size $T$, which defines the non-zero entries for the matrix-vector multiplication. Everything outside of this region is neglected when calculating $\mathcal{A}\vec{x}$ for $\tilde{\Omega}_i$. In the magnified circle in Figure \ref{fig:param}, an illustration of the Gaussian basis functions is shown, where $h$ is the average distance between the calculation points and $\sigma$ is the standard deviation of the Gaussian distribution. When $\sigma$ is large compared to $h$, the matrix $\mathcal{A}$ becomes ill-conditioned. In other words, the more global the basis functions are, the slower the convergence. On the other hand, if the locality was perfect and changing the coefficient $\lambda_j$ in Equation \ref{eq:rbf2} affects only the data at that point $f_j$, the linear system $\mathcal{A}$ would be diagonal, \textit{i.e.} perfectly conditioned, but would suffer from poor approximation properties.

The following function by Franke is a standard test function for 2D scattered data fitting \cite{franke1982}:
\begin{eqnarray}
f(x,y)&=&\frac{3}{4}e^{-\frac{1}{4}((9x-2)^2+(9y-2)^2)}+\frac{3}{4}e^{-\frac{1}{49}(9x+1)^2-\frac{1}{10}(9y+1)^2}\\
&+&\frac{1}{2}e^{-\frac{1}{4}((9x-7)^2+(9y-3)^2)}-\frac{1}{5}e^{-(9x-4)^2-(9y-7)^2}\nonumber
\end{eqnarray}

\noindent We will use this function as a test case throughout the present paper. The points are distributed on a square lattice on the domain $[0,1]^2$. We also used quasi-scattered data, by shifting the position using a random number between $0$ and $h/2$, where $h$ is the average distance between points.

\subsection{Parametric study of convergence rate}

We first investigate the rate of convergence of the present method for different overlap ratio of the Gaussian basis function, $h/\sigma$. The size of the domain is set from $0$ to $1$ and $\sigma=0.01$, which results in $N=(2/h+1)^2=10,201$ points for $h/\sigma=1.0$. The size of the subdomains $B$ is set to $5\sigma$, the overlapping subdomains are of size $D=1.9\times B$, and the truncated domain for the mat-vec products has size $T=B+4\sigma$ ($T=B+6\sigma$ for $h/\sigma=0.8$). These numbers are later shown to be the optimal set of parameters that result in the fastest calculation time. Upon execution of the code, we passed the following command line options to \petsc:

\begin{verbatim}
-pc_type asm -sub_pc_type lu -sub_mat_type dense
-vecscatter_alltoall -ksp_rtol 1e-13 -ksp_monitor -log_summary
\end{verbatim}

\begin{figure}
  \centering
  \includegraphics[keepaspectratio=true,width=0.5\textwidth]{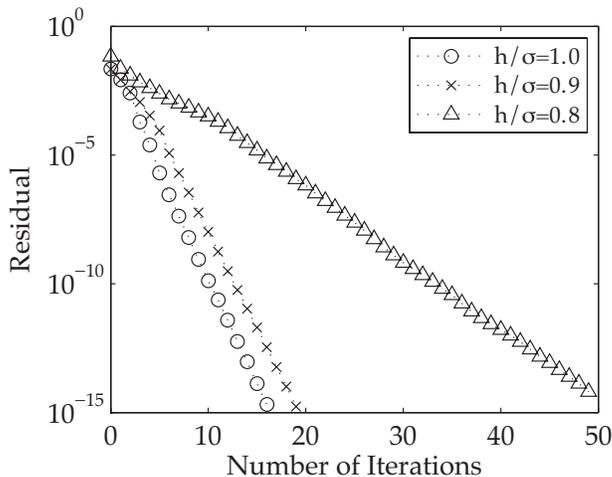}
  \caption{Convergence rate of the \gmres iteration}
  \label{fig:residual}
\end{figure}

\begin{figure}
  \centering
  \subfloat[Calculation time]{
  \includegraphics[keepaspectratio=true,width=0.48\textwidth]{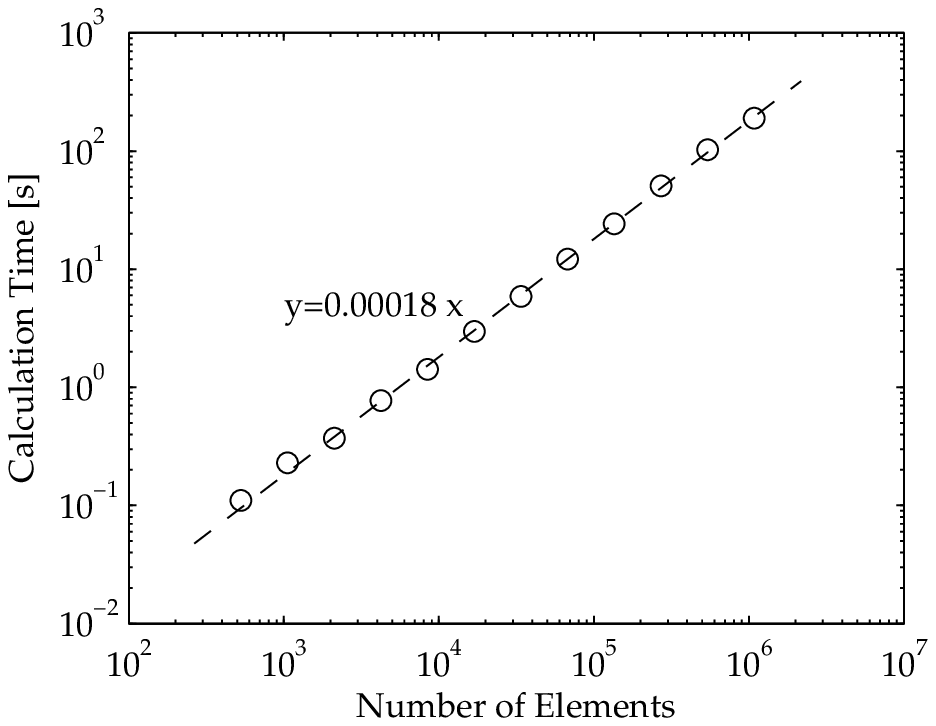}\label{fig:ordern}}
  \subfloat[Memory usage]{
  \includegraphics[keepaspectratio=true,width=0.48\textwidth]{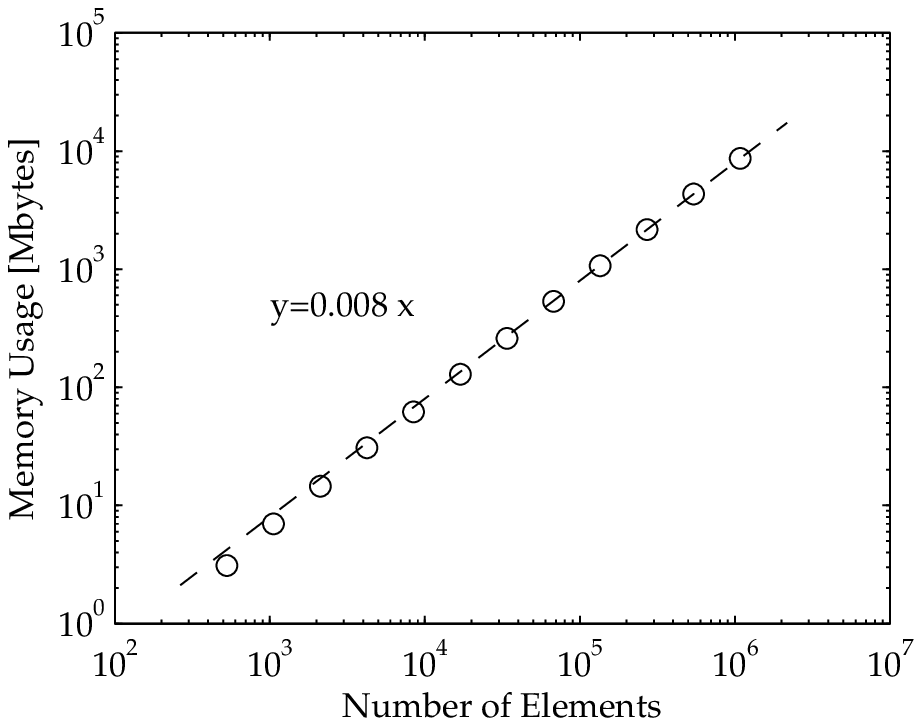}\label{fig:storage}}
  \caption{Scaling of the calculation time and memory usage against the problem size.}
  \label{fig:scaling}
\end{figure}

The $l_2$-norm of the residual $||r||_2$ of the \gmres iterations is plotted against the number of iterations in Figure \ref{fig:residual}. All cases converged to $||r||_2=10^{-15}$, but for $h/\sigma=0.8$ the number of iterations required is significantly larger. This is due to the system $\mathcal{A}$ becoming more ill-conditioned. Nevertheless, the present method converges within 20 iterations for $h/\sigma\ge0.9$. Furthermore, tests using larger $N$ have shown that this convergence rate does not change with the problem size. This can be indirectly observed in the $\mathcal{O}(N)$ behavior of our method shown in Figure \ref{fig:scaling}\subref{fig:ordern}. Since each iteration takes $\mathcal{O}(N)$ time, the number of iterations must remain constant for the entire algorithm to scale as $\mathcal{O}(N)$. The dashed line in Figure \ref{fig:scaling}\subref{fig:ordern} is the function $y=0.00018 x$, which is drawn in order to quantify the scaling exponent. The coefficient $0.00018$ depends on the hardware, and implementation. The memory usage is plotted against the number of elements in Figure \ref{fig:scaling}\subref{fig:storage}. The dashed line is the function $y=0.008 x$, which is drawn in order to quantify the scaling exponent. The storage requirements of the present method also scales as $\mathcal{O}(N)$.

\begin{figure}
  \centering
  \subfloat[$h/\sigma=1.0$]{
  \includegraphics[keepaspectratio=true,width=0.45\textwidth]{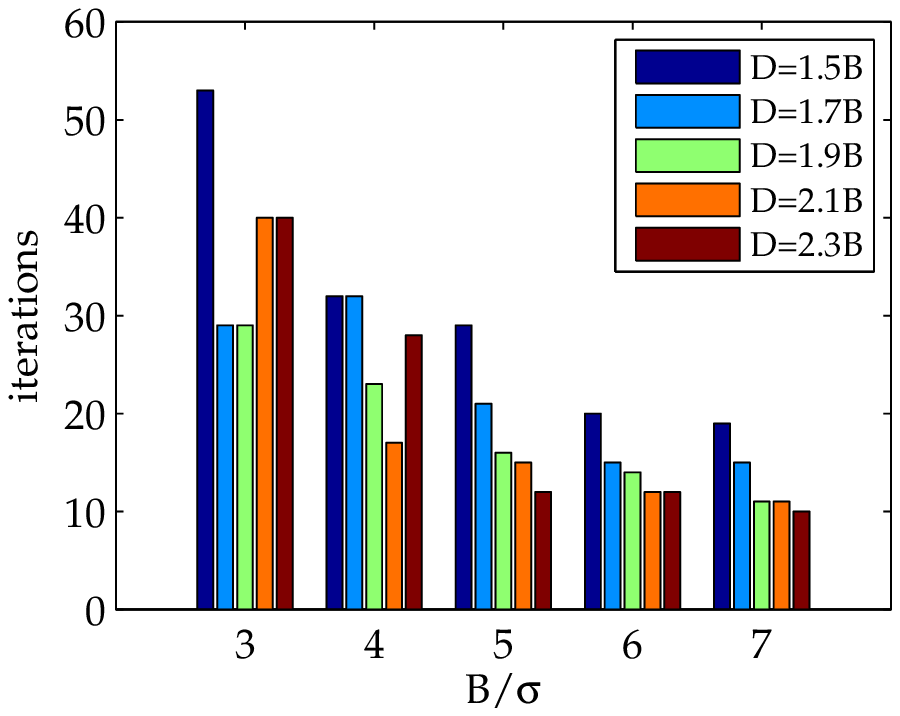}\label{fig:bar_it_10}}
  \subfloat[$h/\sigma=0.9$]{
  \includegraphics[keepaspectratio=true,width=0.45\textwidth]{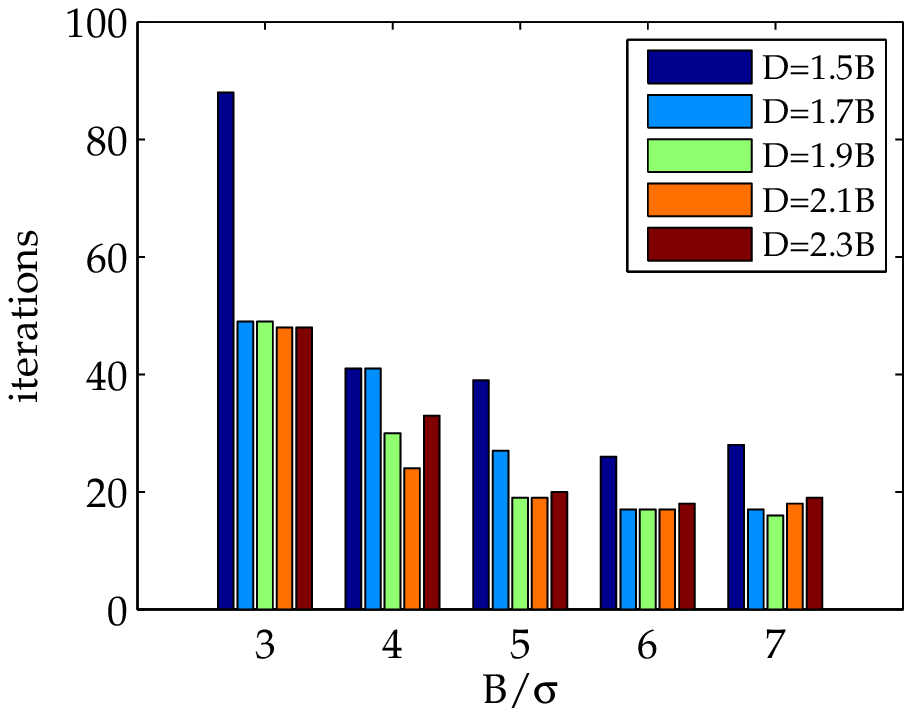}\label{fig:bar_it_9}}\\
  \subfloat[$h/\sigma=0.8$]{
  \includegraphics[keepaspectratio=true,width=0.45\textwidth]{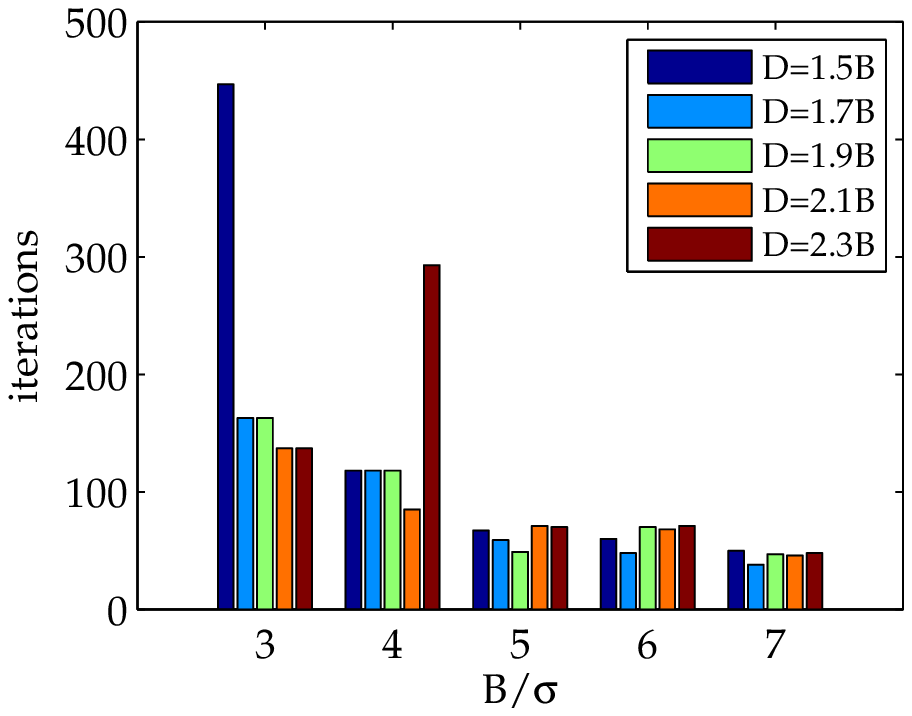}\label{fig:bar_it_8}}
  \caption{Number of iterations for different $B$ and $D$ (data points are on a square lattice).}
  \label{fig:bar_it}
\end{figure}

The rate of convergence does not change with the problem size $N$ as long as the ratio between $h/\sigma$, $B/\sigma$, and $D/\sigma$ remain constant. The maximum number of $N$ that has been calculated is $50,580,544$, and we discuss this result later in Section \ref{parallel}. Even for this case, the number of iterations it takes to converge to $||r||_2=10^{-15}$ remains exactly the same. This is a unique property of the basis function that we are using. For multiquadric and polyharmonics, some sort of hierarchical method would be required in order to have a constant convergence rate.

We now investigate the effect of changing $B$ and $D$, to optimize the performance of the \rasm. The number of iterations is takes to converge to $||r||_2=10^{-15}$ is shown in Figure \ref{fig:bar_it} for different $B$, $D$ and $h/\sigma$. The different plots are for different values of $h/\sigma$. The standard deviation of the Gaussian is $\sigma=0.01$ for all cases, $B/\sigma$ is changed from $3$ to $7$, while $D$ is changed from $1.5B$ to $2.3B$. These parameters were selected by performing a parameter study for a larger range and determining the optimum values $B=5\sigma$ and $D=1.9B$, and then choosing the values close to it to see how the behavior changes. The optimum value for $T$ is also determined by performing a parameter study for a larger range, but will not be changed here for brevity. The optimum value for $T$ was $B+4\sigma$ for $h/\sigma\ge0.9$ and $B+6\sigma$ for $h/\sigma=0.8$, regardless of $B$ and $D$.

\begin{figure}
  \centering
  \subfloat[$h/\sigma=1.0$]{
  \includegraphics[keepaspectratio=true,width=0.45\textwidth]{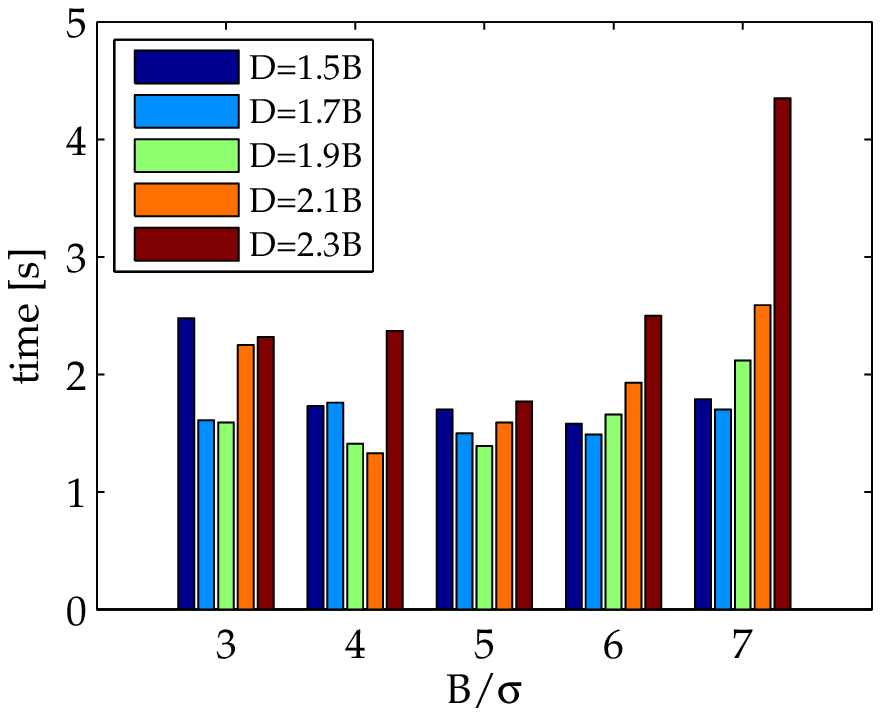}\label{fig:bar_ti_10}}
  \subfloat[$h/\sigma=0.9$]{
  \includegraphics[keepaspectratio=true,width=0.45\textwidth]{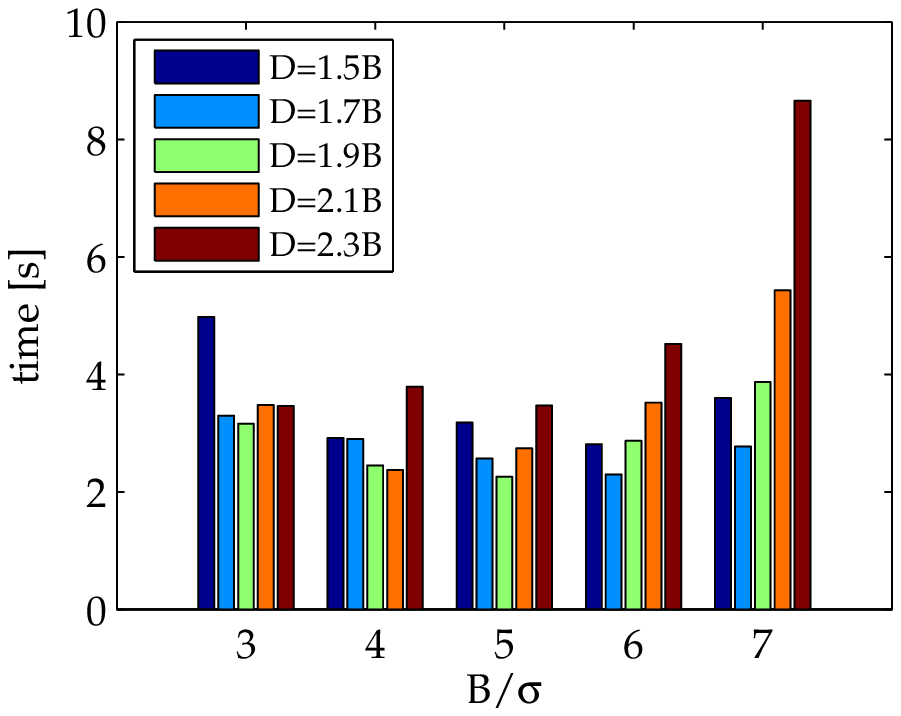}\label{fig:bar_ti_9}}\\
  \subfloat[$h/\sigma=0.8$]{
  \includegraphics[keepaspectratio=true,width=0.45\textwidth]{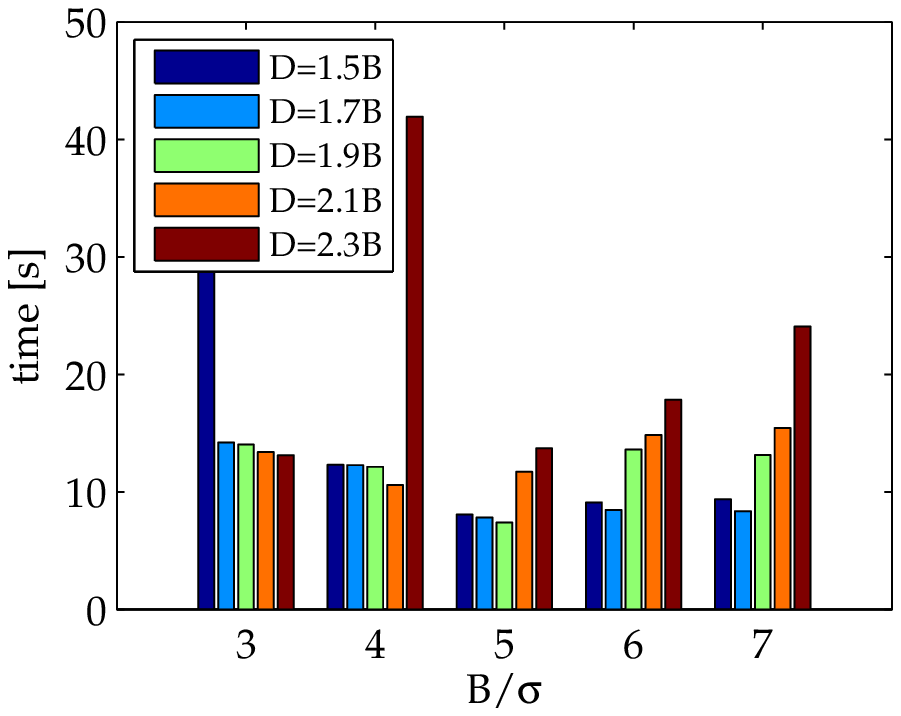}\label{fig:bar_ti_8}}
  \caption{Calculation time for different $B$ and $D$ (data points are on a square lattice).}
  \label{fig:bar_ti}
\end{figure}

It is useful to view Figure \ref{fig:bar_it} in comparison with Figure \ref{fig:bar_ti}, where the calculation time is shown instead of the number of iterations. Keeping $B/\sigma$ constant and increasing $D$ results in the same number of subdomains, but larger overlapping subdomains, so each direct solve for the subdomain takes longer. Since the number of subdomains remains constant, the total calculation time to go through all subdomains increases proportional to each direct solve. On the other hand, a larger overlapping subdomain gives a better convergence rate so fewer iterations will be required to achieve the same accuracy. Thus, there exists an optimum value of $D$ for which the calculation time per iteration is balanced with the required number of iterations to yield the best calculation time. Similarly, increasing $B/\sigma$ results in fewer number of subdomains, but the size of each subdomain will increase along with the time it takes to solve for it. Thus, there exists an optimum value of $B/\sigma$ for which the calculation time per subdomain is balanced with the number of subdomains. It can be seen from Figure \ref{fig:bar_ti} that the optimum set of parameters is $B=5\sigma$ and $D=1.9B$. This optimum set of $B$ and $D$ seems to be consistent throughout the present range of $h/\sigma$.

\section{Scaling of the parallel implementation}
\label{parallel}


There are two common measures for evaluating the performance of parallel codes:  the \emph{strong scaling} and the \emph{weak scaling}. Strong scaling reflects how the solution time varies with the number of processes for a fixed total problem size. Weak scaling, in contrast, shows how the solution time varies with the number of processes for a fixed problem size \emph{per process}. Throughout this paper, we strictly differentiate between processors and processes. Modern processors have multiple cores and can run multiple MPI processes without loss of performance. Thus, the proper unit for parallelization is the MPI process and not the processor.

In the previous section we showed that our algorithm scales as $\mathcal{O}(N)$ in a serial implementation. In this section, we demonstrate that in parallel our algorithm and implementation scale as $\mathcal{O}(N/N_{procs})$ by keeping $N$ constant and changing $N_{procs}$ (strong scaling), and keeping $N/N_{procs}$ constant while changing $N_{procs}$ (weak scaling).

\subsection{Strong scaling}

For the parallel calculations in this section we used the Blue Gene/L machine at the Center for Computational Science (CCS), Boston University \footnote{\href{http://ccs.bu.edu/}{http://ccs.bu.edu/}}, along with the Cray XT4 machine mentioned in the previous section. The Boston University Blue Gene is a single rack system and has 1024 nodes with dual core 32-bit PPC440 processors (700 MHz) and 512 MB of main memory\footnote{See \href{http://scv.bu.edu/computation/bluegene}{http://scv.bu.edu/computation/bluegene}}.

\begin{figure}
  \centering
  \subfloat[$N_{procs}=1$]{
  \includegraphics[keepaspectratio=true,width=0.45\textwidth]{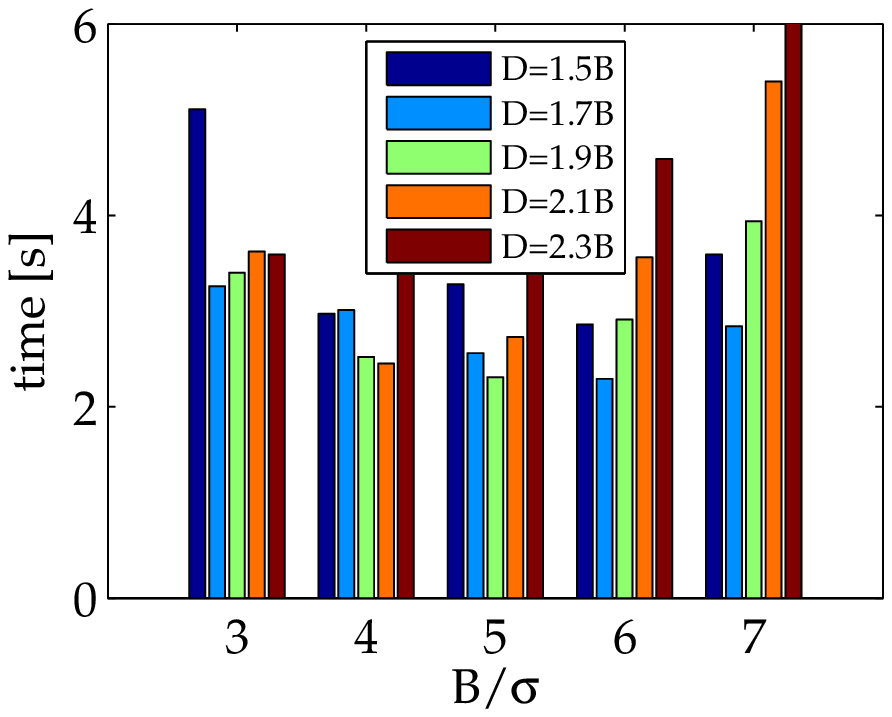}\label{fig:bar_par_1}}
  \subfloat[$N_{procs}=2$]{
  \includegraphics[keepaspectratio=true,width=0.45\textwidth]{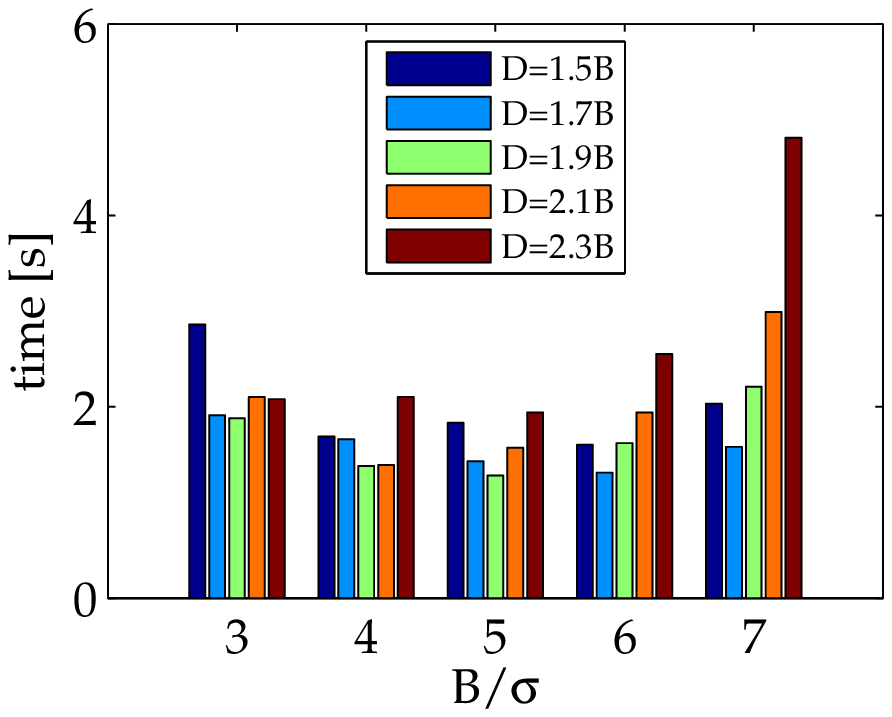}\label{fig:bar_par_2}}\\
  \subfloat[$N_{procs}=4$]{
  \includegraphics[keepaspectratio=true,width=0.45\textwidth]{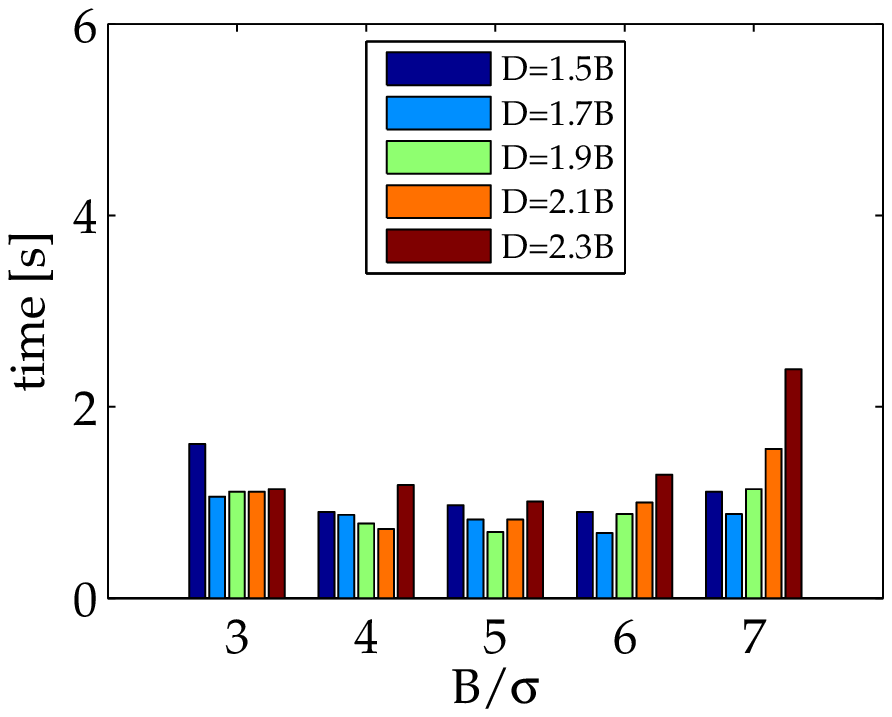}\label{fig:bar_par_4}}
  \subfloat[$N_{procs}=8$]{
  \includegraphics[keepaspectratio=true,width=0.45\textwidth]{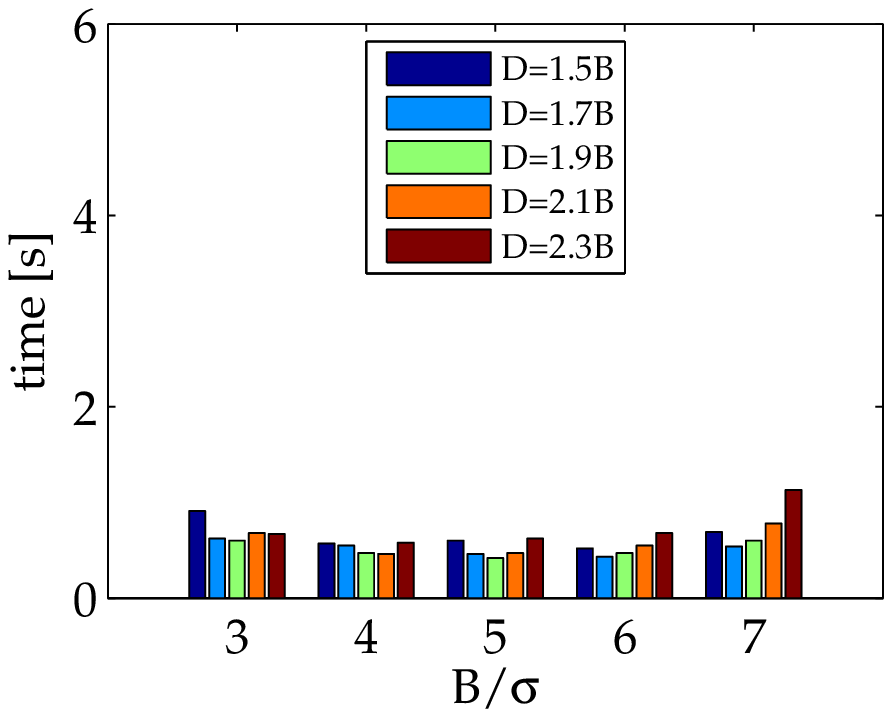}\label{fig:bar_par_8}}
  \caption{Calculation time for different number of processes with varying $B$ and $D$ ($h/\sigma=0.9$, data points are on a square lattice) on Blue Gene/L.}
  \label{fig:bar_par}
\end{figure}

\begin{figure}
  \centering
  \subfloat[$N_{procs}=1$]{
  \includegraphics[keepaspectratio=true,width=0.45\textwidth]{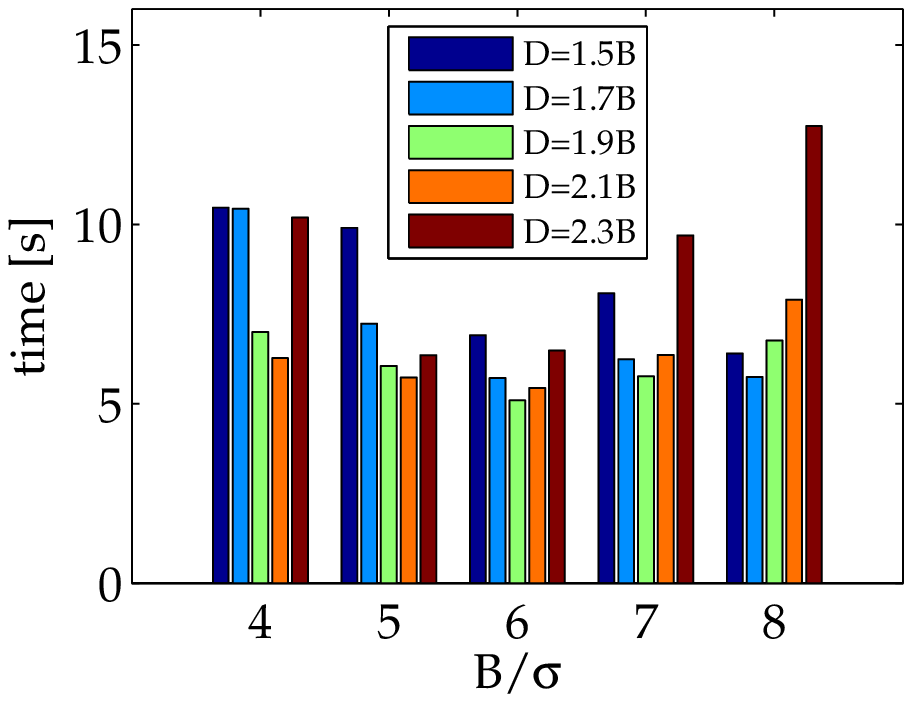}\label{fig:bar_sct_1}}
  \subfloat[$N_{procs}=2$]{
  \includegraphics[keepaspectratio=true,width=0.45\textwidth]{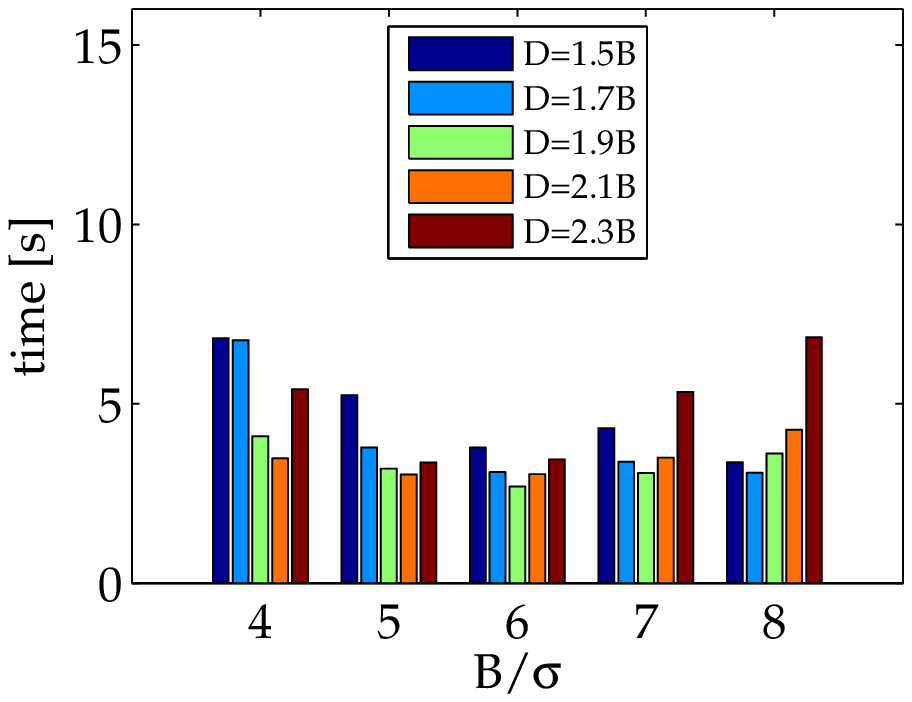}\label{fig:bar_sct_2}}\\
  \subfloat[$N_{procs}=4$]{
  \includegraphics[keepaspectratio=true,width=0.45\textwidth]{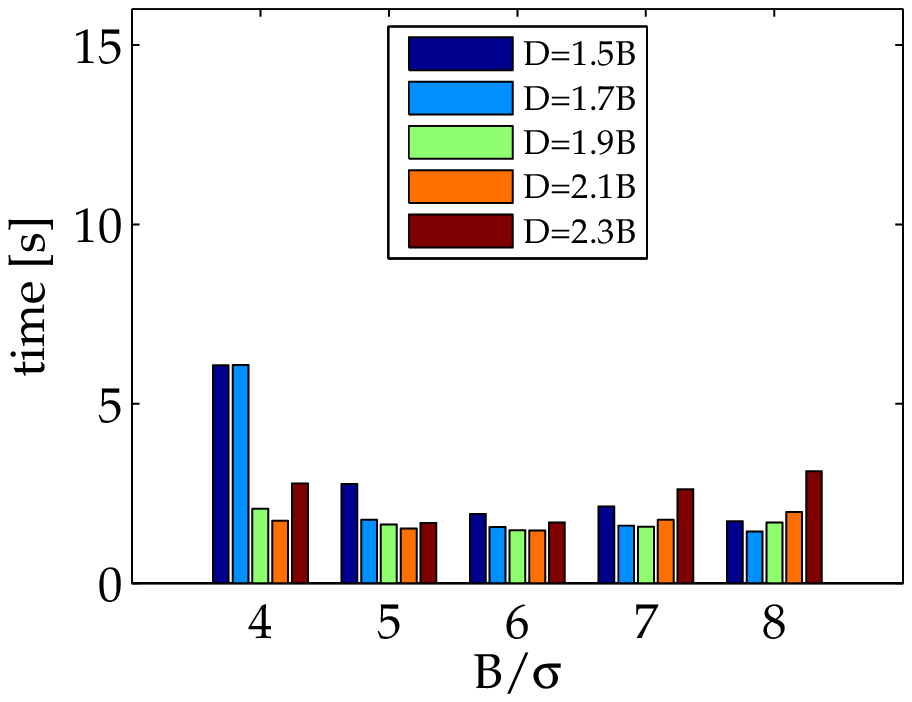}\label{fig:bar_sct_4}}
  \subfloat[$N_{procs}=8$]{
  \includegraphics[keepaspectratio=true,width=0.45\textwidth]{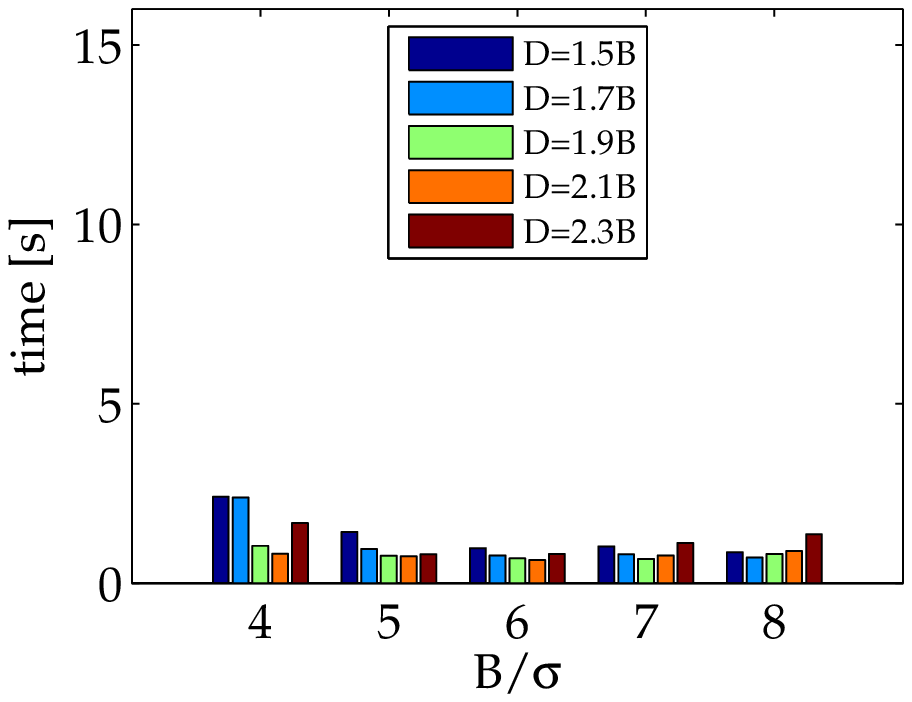}\label{fig:bar_sct_8}}
  \caption{Calculation time for different number of processes with varying $B$ and $D$ ($h/\sigma=0.9$, data points are scattered)  on Blue Gene/L.}
  \label{fig:bar_sct}
\end{figure}

In the present \rbf interpolation method using Gaussian basis functions with small $\sigma$, the problem is highly parallel. This is so because in addition to the fact that the \rasm itself is highly parallel, the matrix vector multiplication for $\mathcal{A}\vec{x}$ can be done using only the information within the influence distance of the Gaussian function ($T$ in Fig.~\ref{fig:param}). Note that this is only possible for rapidly decaying basis functions, where the matrix $\mathcal{A}$ is banded. Fast summation methods for global basis functions will involve more communication and may not scale as well for parallel calculations\,---\,this is the case in \cite{gumerov+duraiswami2007}.

The same calculation as the one shown Figure \ref{fig:bar_ti} was performed in parallel. The results for $1$, $2$, $4$, and $8$ processes are shown in Figure \ref{fig:bar_par}. Note that the results in Figure \ref{fig:bar_par}\subref{fig:bar_par_1} are the same results as the one in Figure \ref{fig:bar_ti}\subref{fig:bar_ti_9}, but the axis has been rescaled. The axes of the four plots in Figure \ref{fig:bar_par} have been aligned so that the speed-up is better represented. The optimum parameter choices of $B=5\sigma$ and $D=1.9B$ do not change when the number of processes is increased. The scaling is not quite what we expect it to be because the problem size $N=(2/h+1)^2=12,544$ is rather small, causing the otherwise insignificant serial parts of the program to degrade the scalability.

The results shown so far are for equally spaced data points on a square lattice. We now consider the effect of having scattered data points. The same calculation as in Figure \ref{fig:bar_par} is performed for the same number of scattered data points. The results are shown in Figure \ref{fig:bar_sct}. For scattered data, the optimum size of the $B$, $D$, and $T$ domain sketched in Figure \ref{fig:param} changed slightly to $B=6\sigma$, $D=1.9B$, and $T=B+6\sigma$ for $h/\sigma=0.9$. Compared to the case using a square lattice, the optimum value for $B$ increases by $\sigma$ and $T$ increases by $2\sigma$, while $D/B$ remains the same. The domain must be slightly larger than the square lattice case, because the random positioning of data points results in a slightly more ill-conditioned system. The calculation time also increases due to the number of iterations increasing. However, in our target application we use \rbf interpolation to interpolate from scattered points onto an equally-spaced lattice distribution. Therefore, the performance of the interpolation onto scattered data points is not of primary importance. We will note that the difference is small, and proceed with our analysis for the parallel performance using the data points on a square lattice.

\begin{figure}
  \centering
  \subfloat[Strong scaling]{
  \includegraphics[keepaspectratio=true,width=0.48\textwidth]{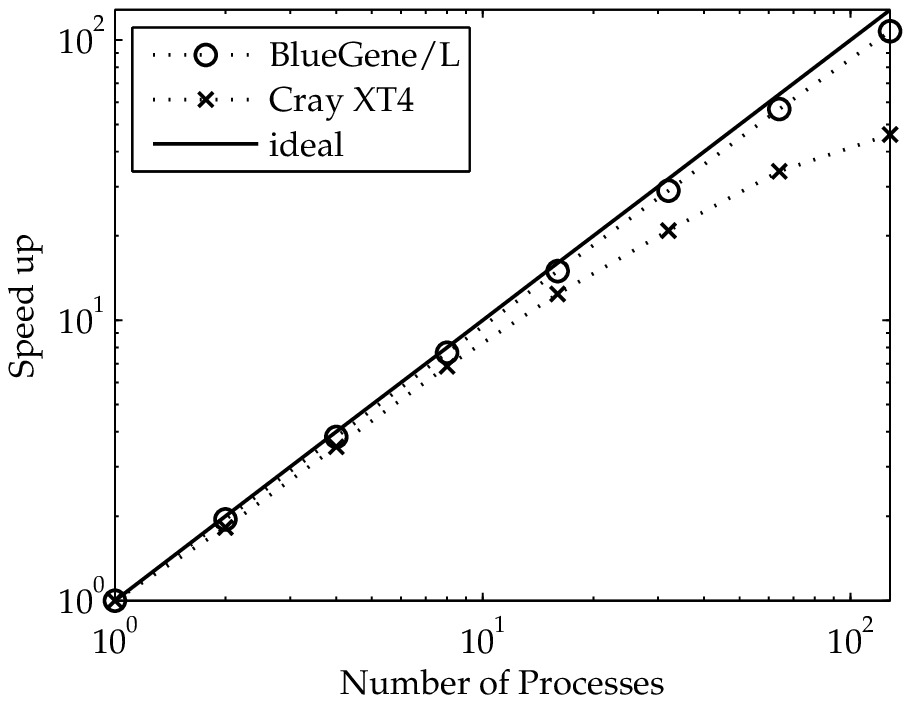}\label{fig:strong}}
  \subfloat[Memory usage]{
  \includegraphics[keepaspectratio=true,width=0.48\textwidth]{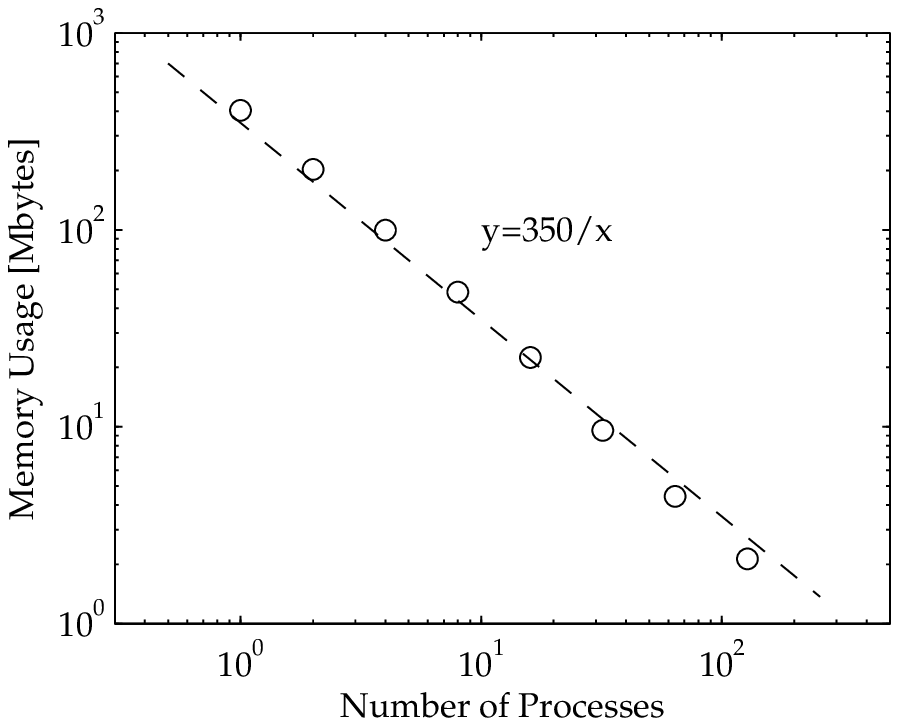}\label{fig:storage_strong}}
  \caption{Strong scaling and memory usage for different number of processes.}
  \label{fig:strongs}
\end{figure}

We now select the optimum parameter $B=5\sigma$ and $D=1.9B$ and test the speed-up for a larger number of processes. The calculation is performed on the Cray XT4 used above, and also a Blue Gene/L. The strong scaling results are shown in Figure \ref{fig:strongs}\subref{fig:strong}. Preliminary calculations have shown that smaller $h/\sigma$ gives slightly better scaling, but only the results for $h/\sigma=0.9$ will be shown here. Also, $\sigma$ is set to $0.005$ for which case $N=49,729$. The memory limitation on the Blue Gene/L $512\ MB/node$ prevented the strong scaling tests for larger calculations. The parallel efficiency using for 128 processes is $84\%$ for the Blue Gene/L and $36\%$ for the Cray XT4, using this test problem.

The memory usage for the strong scaling tests in shown in Figure \ref{fig:strongs}\subref{fig:storage_strong}. The dashed line is the function $y=350/x$. It can be seen that the memory usage decreases inversely proportional (or even better) to the number of processes $N_{procs}$. Thus, it is safe to say that we have a storage requirement of $\mathcal{O}(N/N_{procs})$. For the weak scaling, the $\mathcal{O}(N/N_{procs})$ storage of our method allows the calculation of large-scale problems. This will be shown in the next subsection.

\subsection{Weak scaling}

\begin{figure}
  \centering
  \subfloat[Parallel Efficiency]{
  \includegraphics[keepaspectratio=true,width=0.48\textwidth]{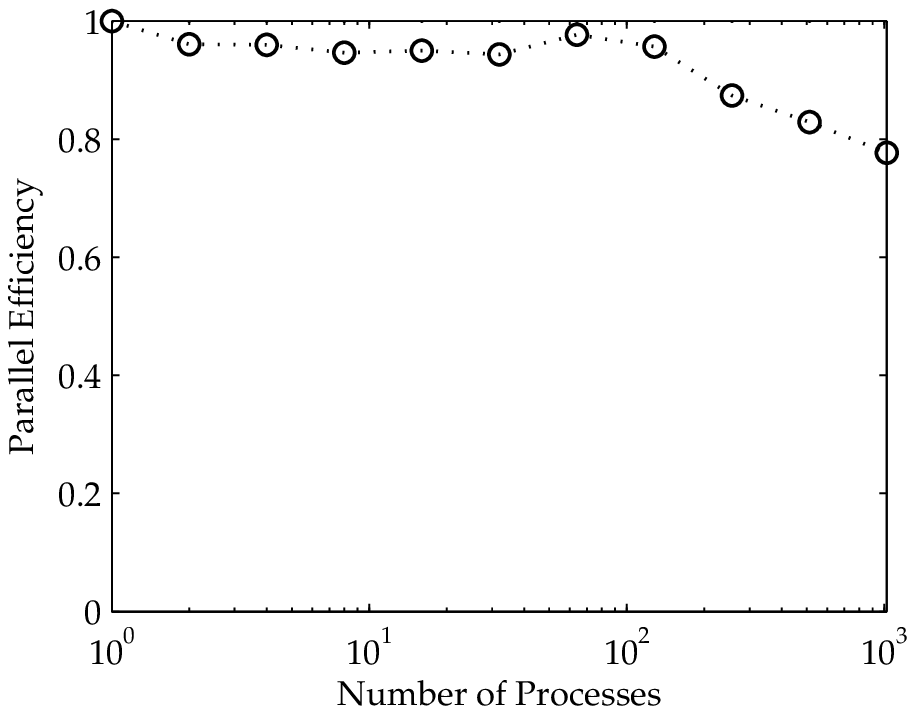}\label{fig:efficiency}}
  \subfloat[Breakdown of calculation time]{
  \includegraphics[keepaspectratio=true,width=0.48\textwidth]{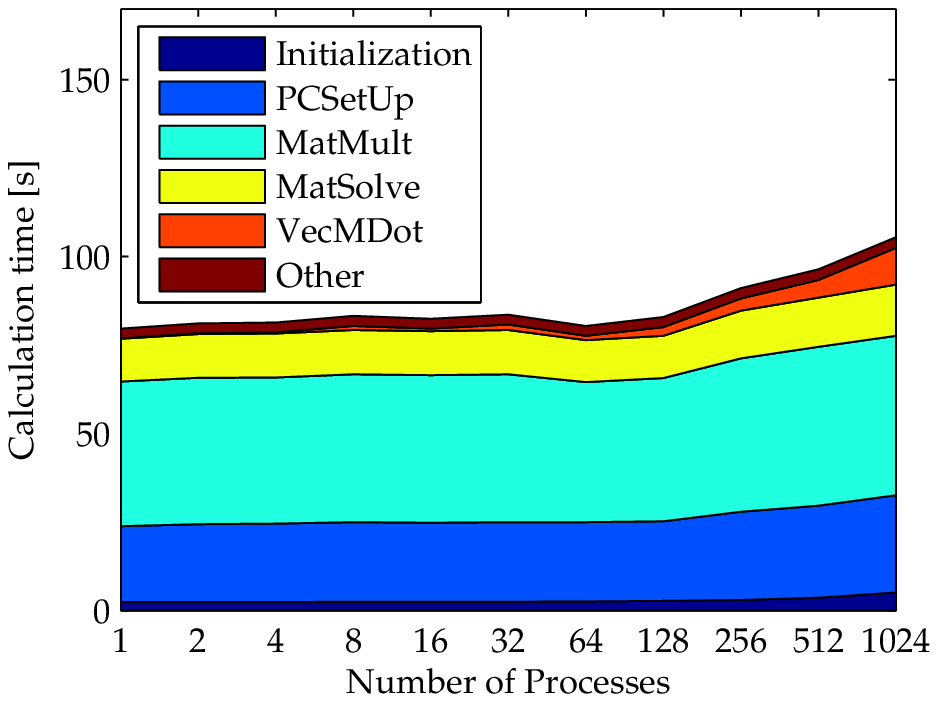}\label{fig:breakdown}}
  \caption{Weak scaling on the Blue Gene/L.}
  \label{fig:weak}
\end{figure}

The weak scaling on Blue Gene/L is shown in Figure \ref{fig:weak}\subref{fig:efficiency}. The parallel efficiency is the speed-up divided by the number of processes. Parallel efficiency equal to unity means that the parallel implementation scales perfectly with the serial implementation. The number of calculation points grows from $49,729$ (1 process) to $50,580,544$ ($1024$ processes). Even under the contraints of $512\ MB/node$, the $\mathcal{O}(N/N_{procs})$ storage of our code enabled the calculation of large problems. The weak scaling is almost perfect until $128$ processes and gradually declines after that. The parallel efficiency for $1024$ processes is $78\%$.

The breakdown of the calculation time for the weak scaling test is shown in Figure \ref{fig:weak}\subref{fig:breakdown}. ``Initialization'' corresponds to the allocation and initialization of all variables, ``PCSetUp'' is the setup of the \rasm preconditioner, ``MatMult'' is the matrix-vector multiplication, ``MatSolve'' is the matrix solver for the subdomains, ``VecMDot'' is the dot product of vectors, and ``Other'' is the total of everything else. The total calculation time is the sum of all colors, which is around 80 seconds for 1 process and around 106 seconds for 1024 processes. The curve in Figure \ref{fig:weak}\subref{fig:efficiency} is a vertical flip of the curve at the top of Figure \ref{fig:weak}\subref{fig:breakdown}. Fig.~\ref{fig:weak}\subref{fig:breakdown} gives a clear view of which part of the code is degrading the weak scaling performance. It is mainly the communication time in ``PCSetUp'' and ``VecMDot'' that is increasing the overall calculation time. Further tuning of these communication routines could allow the present code to scale up to tens of thousands of processors. However, such tuning is too hardware-specific to be discussed in the present paper.

\section{Conclusions}
\label{conclusions}



This work positions itself among a series of recent efforts to provide fast algorithms for the solution of radial basis function (\rbf) interpolation problems.  The subject has gained much momentum in recent years, due to growing enthusiasm over meshfree methods, reflected in new books, such as \cite{fasshauer2007}, and various international conferences dedicated to the topic.  Meshfree methods are attractive due to the infamous hurdle of mesh generation and maintenance for many applications, especially those involving complex or moving geometries.  Moreover, \rbf interpolation is an important technique for approximation of unknown functions based on unordered data.  Applications include solution of partial differential equations \cite{kansa1990a,kansa1990b}, reconstruction of surfaces for design of medical implants based on imaging \cite{CarrFrightBeatson1997}, interpolation of geophysical data \cite{billings+beatson+newsam2002}, modeling of groundwater contaminant transport \cite{LiChenPepper2003}, to name just a few.

In view of the multitude of important applications, the effort to offer computational efficiency has been strong.  There have persistently been two challenges for this effort to succeed:  the need to solve a dense and ill-conditioned linear system, and the cost of evaluating the \rbf interpolant once a solution is found.  This last challenge impacts on the first when using iterative methods of solution, as the internal iterations require an \rbf evaluation (a dense matrix-vector multiplication).  Until recently, the only way to address these challenges was to opt for using basis functions of compact support \cite{wendland1995,buhmann1995b}.  The development of such compact support bases had a significant influence.  It is known, however, that the approximation properties of compact support bases are inferior to global bases \cite[p.150]{buhmann2003}.  For this reason, several authors have continued to investigate ways to provide fast \rbf interpolation with global bases.

Our contribution consists of an \rbf interpolation algorithm with \bigON\ complexity, and \bigON\ storage.  Moreover, we have formulated the algorithm \emph{in parallel}.  As far as we could find in the published literature, the only other parallel \rbf method was presented in \cite{IngberChenTanski2004}.  That work used polyharmonic basis functions, and parallel efficiency was rather modest; the largest problem reportedly solved involved $20,000$ interpolation points.

Two other \rbf algorithms of note are the one developed by Beatson and others \cite{beatson.et.al2000} and the accelerated version by Gumerov and Duraiswami \cite{gumerov+duraiswami2007} of a method developed in \cite{faul+goodsell+powell2005}.  Both of these were implemented for serial computations only (as far as we are aware), and both have $\mathcal{O}(N \log N)$ complexity. In the first case, a multiplicative domain decomposition approach was used with polyharmonic basis functions.  The numerical demonstrations there represent the largest problem size we could find reported, with 5 million interpolation points in 2D.  In the second case, a preconditioner based on approximate cardinal functions is applied \cite{beatson1993,beatsonETal1999}, and both multi-quadric and polyharmonic bases are used.  There, the largest calculation reported interpolated 1 million data points in 2D. As the work in \cite{gumerov+duraiswami2007} is very recent, the hardware used is comparable to the one used in our results.  The reported timing for 1 million points was over 6000 seconds (in one 3.2 GHz processor).  Although the basis function used is different, as well as the method of solution, a casual comparison can be made with our results.  As shown in Figure \ref{fig:scaling}\subref{fig:ordern}, we have timed a solution with 1 million points at about 200 seconds (with a slower 2.3 GHz processor).  Moreover, the accuracy we imposed in our experiment was higher by 7 orders of magnitude.

In light of the publications cited above, we are persuaded that the present work positions itself as the fastest and most efficient \rbf algorithm to date.  We aim to multiply the impact of our algorithm by realizing an implementation that is based on a well-tuned, well-supported parallel library for scientific computing, \petsc, and by offering the software free of charge, and in the open model.
 
\medskip

In summary, our algorithm and its implementation have demonstrated good asymptotic behavior in the following aspects:
\begin{itemize}
\item[$\triangleright$] Complexity: $\mathcal{O}(N)$ for the observed range of $N=10^3-10^6$ \vspace{-0.5em}
\item[$\triangleright$] Storage: $\mathcal{O}(N)$ for the observed range of $N=10^3-10^6$ \vspace{-0.5em}
\item[$\triangleright$] Convergence Rate: $\mathcal{O}(1)$ for the observed range of $N$ (less than $20$ iterations for most cases) \vspace{-0.5em}
\item[$\triangleright$] Strong scaling: parallel efficiency of $84\%$ on $128$ processes for $N=49,729$ \vspace{-0.5em}
\item[$\triangleright$] Weak scaling: parallel efficiency of $78\%$ on $1024$ processes for $N/N_{procs}\approx49,729$ \vspace{-0.5em}
\item[$\triangleright$] Parallel storage: $\mathcal{O}(N/N_{procs})$ for the range $N_{procs}=1-128$ 
\end{itemize}

The largest \rbf interpolation problem we have calculated so far had over 50 million data points, and this calculation was timed at $106$ seconds ($19$ iterations, for an accuracy of $10^{-15}$) using 1024 processors on a Blue Gene/L supercomputer. Considering the fact that the processors of the Blue Gene/L are significantly slower than the high-end CPUs available today (about 5 times) and have only $512\ MB/node$ of memory, the performance is quite remarkable.

\bigskip

For information about downloading the parallel software, which has been dubbed \textsc{p}et\textsc{rbf} (for `parallel, extensible toolkit for \rbf interpolation),  please visit the website at \href{http://barbagroup.bu.edu/}{http://barbagroup.bu.edu/}.

\section*{Acknowledgments}
Computing time provided by the \href{http://ccs.bu.edu/}{Center for Computational Science (CCS)} of Boston University, on BlueGene/L, and the UK National Supercomputing Service, on the \href{http://ccs.bu.edu/}{\hector} supercomputer. LAB acknowledges partial support from EPSRC under grant contract EP/E033083/1, and from Boston University College of Engineering.


\bibliographystyle{plain}
\bibliography{vortexmeth,vortexflow,scicomp,meshfree,FastMethods}

\begin{thebibliography}{10}

\bibitem{atluri1998}
S.~N. Atluri and T.~Zhu.
\newblock A new meshless local {P}etrov-{G}alerkin ({MLPG}) approach in
  computational mechanics.
\newblock {\em Computational Mechanics}, 22:117--127, 1998.

\bibitem{babuska1997}
I.~Babuska and J.~M. Melenk.
\newblock The partition of unity method.
\newblock {\em International Journal for Numerical Methods in Engineering},
  40:727--758, 1997.

\bibitem{petsc-manual}
S.~Balay, K.~Buschelman, W.~D. Gropp, D.~Kaushik, M.~Knepley,
  L.~Curfman-McInnes, B.~F. Smith, and H.~Zhang.
\newblock {PETSc} {U}ser's {M}anual.
\newblock Technical Report ANL-95/11 - Revision 2.1.5, Argonne National
  Laboratory, 2002.

\bibitem{barba2005}
L.~A. Barba.
\newblock Computing high-{R}eynolds number vortical flows: a highly accurate
  method with a fully meshless formulation.
\newblock In {\em Parallel Computational Fluid Dynamics\,---\,Multidisciplinary
  Applications}, pages 305--312. Elsevier B.~V., 2005.

\bibitem{barba+leonard2007}
L.~A. Barba and A.~Leonard.
\newblock Emergence and evolution of tripole vortices from net-circulation
  initial conditions.
\newblock {\em Phys.\ Fluids}, 19(1):017101, 2007.

\bibitem{barbaETal2005}
L.~A. Barba, A.~Leonard, and C.~B. Allen.
\newblock Advances in viscous vortex methods -- meshless spatial adaption based
  on radial basis function interpolation.
\newblock {\em Int.\ J. Num.\ Meth.\ Fluids}, 47(5):387--421, 2005.

\bibitem{BarbaRossi2009}
L.~A. Barba and L.~F. Rossi.
\newblock Global field interpolation for particle methods.
\newblock \emph{J.\ Comput.\ Phys.}, 2009.
\newblock To appear.

\bibitem{beatsonETal1999}
R.~K. Beatson, J.~B. Cherrie, and C.~T. Mouat.
\newblock Fast fitting of radial basis functions: {M}ethods based on
  preconditioned {GMRES} iteration.
\newblock {\em Adv.\ Comp.\ Math.}, 11(2--3):253--270, November 1999.

\bibitem{beatson.et.al2000}
R.~K. Beatson, W.~A. Light, and S.~Billings.
\newblock Fast solution of the radial basis function interpolation equations:
  {D}omain decomposition methods.
\newblock {\em SIAM J. Sci.\ Comput.}, 22(5):1717--1740, 2000.

\bibitem{beatson+newsam1992}
R.~K. Beatson and G.~N. Newsam.
\newblock Fast evaluation of radial basis functions: {I}.
\newblock {\em Comp.\ Math.\ Applic.}, 24(12):7--19, 1992.

\bibitem{beatson1993}
R.~K. Beatson and M.~J.~D. Powell.
\newblock An iterative method for thin plate spline interpolation that employs
  approximations to {L}agrange functions.
\newblock In D.~F. Griffiths and G.~A. Watson, editors, {\em Numerical
  Analysis}, pages 17--39. Longman Sci. Tech., 1993.

\bibitem{BelytschkoETal94}
Ted Belytschko, Y.~Y. Lu, and L.~Gu.
\newblock Element-free {G}alerkin methods.
\newblock {\em Int.\ J. Num.\ Meth.\ Eng.}, 37(2):229--256, 1994.

\bibitem{billings+beatson+newsam2002}
S.~D. Billings, R.~K. Beatson, and G.~N. Newsam.
\newblock Interpolaton of geophysical data using continuous global surfaces.
\newblock {\em Geophysics}, 67(6):1810--1822, 2002.

\bibitem{buhmann1995b}
M.~D. Buhmann.
\newblock Radial functions on compact support.
\newblock {\em Proc.\ Edinburgh Math\. Soc.}, 41(33--46), 1998.

\bibitem{buhmann2003}
M.~D. Buhmann.
\newblock {\em Radial Basis Functions. Theory and Implementations}.
\newblock Cambridge University Press, 2003.

\bibitem{cai1997}
X.-C. Cai and M.~Sarkis.
\newblock A restricted additive schwarz preconditioner for general sparse
  linear systems.
\newblock {\em SIAM Journal on Scientific Computing}, 21:792--797, 1997.

\bibitem{CarrFrightBeatson1997}
J.~C. Carr, W.~R. Fright, and R.~K. Beatson.
\newblock Surface interpolation with radial basis functions for medical
  imaging.
\newblock {\em IEEE Trans.\ on Medical Imaging}, 16(1):96--107, 1997.

\bibitem{chen2002a}
W.~Chen.
\newblock New {RBF} collocation schemes and kernel {RBF}s with applications.
\newblock {\em Lecture Notes in Computational Science and Engineering},
  26:75--86, 2002.

\bibitem{cherrie+beatson+newsam2002}
J.~B. Cherrie, R.~K. Beatson, and G.~N. Newsam.
\newblock Fast evaluation of radial basis functions: {M}ethods for generalized
  multiquadrics in ${R}^n$.
\newblock {\em SIAM J. Sci.\ Comput.}, 23(5):1549--1571, 2002.

\bibitem{DuarteOden96}
C.~A. Duarte and J.~Tinsley Oden.
\newblock An $h$-$p$ adaptive method using clouds.
\newblock {\em Comp.\ Meth.\ Appl.\ Mech.\ Engrg.}, 139(1--4):237--262, 1996.

\bibitem{efstathiou2003}
E.~Efstathiou and M.~J. Gander.
\newblock Why restricted additive {S}chwarz converges faster than additive
  {S}chwarz.
\newblock {\em BIT Numerical Mathematics}, 43:945--959, 2003.

\bibitem{fasshauer1997}
G.~E. Fasshauer.
\newblock Solving partial differential equations by collocation with radial
  basis functions.
\newblock In Le~Mehaute A., Rabut C., and Schumaker~L. L., editors, {\em
  Surface Fitting and Multiresolution Methods}, pages 131--138. Vanderbilt
  University Press, 1997.

\bibitem{fasshauer2007}
Gregory~E. Fasshauer.
\newblock {\em Meshfree Approximation Methods with {MATLAB}}.
\newblock World Scientific, 2007.

\bibitem{faul+goodsell+powell2005}
A.~C. Faul, G.~Goodsell, and M.~J.~D. Powell.
\newblock A {K}rylov subspace algorithm for multiquadric interpolation in many
  dimensions.
\newblock {\em IMA J. Numer.\ Anal.}, 25:1--24, 2005.

\bibitem{franke1982}
R.~Franke.
\newblock Scattered data interpolation: {T}ests of some methods.
\newblock {\em Math.\ Comp.}, 38(157):181--200, 1982.

\bibitem{greengard+strain1991}
L.~Greengard and J~Strain.
\newblock The fast {Gauss} transform.
\newblock {\em SIAM Journal on Scientific and Statistical Computing},
  12(1):79--94, January 1991.

\bibitem{gumerov2005}
N.~A. Gumerov and R.~Duraiswami.
\newblock Fast multipole method for the biharmonic equation.
\newblock Technical Report UMIACS-TR-2005-29, Technical Reports of the Computer
  Science Department, University of Maryland, 2005.

\bibitem{gumerov+duraiswami2007}
N.~A. Gumerov and R.~Duraiswami.
\newblock Fast radial basis function interpolation via preconditioned {K}rylov
  iteration.
\newblock {\em SIAM J. Sci.\ Comput.}, 29(5):1876--1899, 2007.

\bibitem{hardy1990}
R.~L. Hardy.
\newblock Theory and applicatoins of the multiquadric-biharmonic method.
\newblock {\em Computers and Mathematics with Applications}, 19(8/9):163--208,
  1990.

\bibitem{HughesETal2005}
T.~J.~R. Hughes, J.~A. Cottrell, and Y.~Bazilevs.
\newblock Isogeometric analysis: {CAD}, finite elements, {NURBS}, exact
  geometry and mesh refinement.
\newblock {\em Comput.\ Methods Appl.\ Mech.\ Engrg.}, 194(39--41):4135--4195,
  2005.

\bibitem{IngberChenTanski2004}
M.~S. Ingber, C.~S. Chen, and J.~A. Tanski.
\newblock A mesh free approach using radial basis functions and parallel domain
  decomposition for solving three-dimensional diffusion equations.
\newblock {\em Int.\ J. Num.\ Meth.\ Eng.}, 60:2183--2201, 2004.

\bibitem{kansa1990a}
E.~J. Kansa.
\newblock Multiquadrics ---{A} scattered data approximation scheme with
  applications to computational fluid-dynamics, {I}. {S}urface approximations
  and partial derivative estimates.
\newblock {\em Computers Math.\ Applic.}, 19(8/9):127--145, 1990.

\bibitem{kansa1990b}
E.~J. Kansa.
\newblock Multiquadrics ---{A} scattered data approximation scheme with
  applications to computational fluid-dynamics, {II}. {S}olutions to parabolic,
  hyperbolic and elliptic partial differential equations.
\newblock {\em Computers Math.\ Applic.}, 19(8/9):147--161, 1990.

\bibitem{kansa+hon2000}
E.~J. Kansa and Y.~C. Hon.
\newblock Circumventing the ill-conditioning problem with multiquadric radial
  basis functions: {A}pplications to elliptic partial differential equations.
\newblock {\em Comp.\ Math.\ Appl.}, 39:123--137, 2000.

\bibitem{leonard1980}
A.~Leonard.
\newblock Vortex methods for flow simulation.
\newblock {\em J. Comp.\ Phys.}, 37:289--335, 1980.

\bibitem{LiChenPepper2003}
J.~Li, Y.~Chen, and D.~Pepper.
\newblock Radial basis function method for 1-{D} and 2-{D} groundwater
  contaminant transport modeling.
\newblock {\em Comput.\ Mech.}, 23:10--15, 2003.

\bibitem{li2004}
J.~Li and Y.~C. Hon.
\newblock Domain decomposition for radial basis meshless methods.
\newblock {\em Numerical Methods for Partial Differential Equations},
  20(3):450--462, 2004.

\bibitem{ling2004}
L.~Ling and E.~J. Kansa.
\newblock Preconditioning for radial basis functions with domain decomposition
  methods.
\newblock {\em Mathematical and Computer Modeling}, 40:1413--1427, 2004.

\bibitem{ling2005}
L.~Ling and E.~J. Kansa.
\newblock A least-squares preconditioner for radial basis functions collocation
  methods.
\newblock {\em Advances in Computational Mathematics}, 23:31--54, 2005.

\bibitem{KamLiuETal1995}
Wing~Kam Liu, Sukky Jun, Shaofan Li, Jonathan Adee, and Ted Belytschko.
\newblock Reproducing kernel particle methods for structural dynamics.
\newblock {\em Int.\ J. Num.\ Meth.\ Eng.}, 38(10):1655--1679, 1995.

\bibitem{madych+nelson1990}
W.~R. Madych and S.~A. Nelson.
\newblock Multivariate interpolation and conditionally positive definite
  functions: {II}.
\newblock {\em Math.\ Comp.}, 54(189):211--230, 1990.

\bibitem{mai-duy2003}
N.~Mai-Duy and T.~Tran-Cong.
\newblock Indirect {RBFN} method with thin plate splines for numerical solution
  of differential equations.
\newblock {\em Computer Modeling in Engineering and Sciences}, 4(1):85--102,
  2003.

\bibitem{Micchelli1986}
C.~A. Micchelli.
\newblock Interpolation of scattered data: distance matrices and conditionally
  positive definite functions.
\newblock {\em Constr.\ Approx.}, 2:11--22, 1986.

\bibitem{powell1993}
M.~J.~D. Powell.
\newblock Truncated {L}aurent expansions for the fast evaluation of thin plate
  splines.
\newblock {\em Numerical Algorithms}, 5:99--120, 1993.

\bibitem{roussos2005}
G.~Roussos and B.~J.~C. Baxter.
\newblock Rapid evalutation of radial basis functions.
\newblock {\em Journal of Computational and Applied Mathematics},
  180(1):51--70, 2005.

\bibitem{saad1996}
Y.~Saad.
\newblock {\em Iterative Methods for Sparse Linear Systems}.
\newblock PWS Publishing Co., Boston, 1996.

\bibitem{smith1996}
B.~F. Smith, P.~E. Bjorstad, and W.~D. Gropp.
\newblock {\em Domain Decomposition, Parallel Multilevel Methods for Elliptic
  Partial Differential Equations}.
\newblock Cambridge University Press, 1996.

\bibitem{TorresBarba2009}
C.~E. Torres and L.~A. Barba.
\newblock Fast radial basis function interpolation with {G}aussians by
  localization and iteration.
\newblock {\em J. Comp.\ Phys.}, 228:4976--4999, 2009.

\bibitem{wendland1995}
H.~Wendland.
\newblock Piecewise polynomial, positive definite and compactly supported
  radial basis functions of minimal degree.
\newblock {\em Adv.\ Comp.\ Math.}, 4:389--396, 1995.

\bibitem{wong1999}
S.~M. Wong, Y.~C. Hon, T.~S. Li, and S.~L. Chung.
\newblock Multi-zone decomposition for simulation of time-dependent problems
  using the multiquadric scheme.
\newblock {\em Computers and Mathematics with Applications}, 37:23--43, 1999.

\bibitem{yokota2007}
R.~Yokota, T.~K. Sheel, and S.~Obi.
\newblock Calculation of isotropic turbulence using a pure {L}agrangian vortex
  method.
\newblock {\em J. Comp.\ Phys.}, 226:1589--1606, 2007.

\end{thebibliography}

\end{document}